\documentclass{elsart}
\usepackage{graphicx,amsmath,amsfonts, ecltree, epic, amsmath, amssymb, amsfonts}

\begin{document}
\begin{frontmatter}

\title{Irreversible evolution of a charged spin $1/2$ particle analyzed on the basis of subdynamics theory
}
\author{S.Eh.Shirmovsky}
\ead{shirmov@ifit.phys.dvgu.ru} \address{ Laboratory of Theoretical Nuclear Physics, Far Eastern National University,
8 Sukhanov St., Vladivostok, 690950, Russia}
\date{\today}

\begin{abstract}
The time evolution of a charged spin $1/2$ particle interacting with a magnetic field is analyzed in the framework of the complex spectral theory,
based on Prigogine's principles that provide a rigorous description of irreversibility. A detailed survey of the irreversible relaxation process of the spin $1/2$ particle is carried out on the basis of the subdynamics theory. We obtain the results for the Markovian and the non-Markovian evolution of the charged particle.
 \end{abstract}
\end{frontmatter}

\section{Introduction}
Our experience suggests that the symmetry in the
time is disrupted and the future and the past play different
roles. The world surrounding us has obvious irreversible nature.
However,  Poincare showed that~\cite{poincsre1}  the description of the irreversibility  is impossible on the basis of the classical laws of dynamics, since the latter are reversible in time. 
The equations of quantum mechanics are reversible too. 
The description of the physical world on the basis of
fundamental classical and quantum theories as defined by the laws
of the nature is determined to be time reversible.
Difference between the classical description of the nature 
and those processes in the nature which we observe creates the
conflict situation. Therefore the problem of the description of  irreversible world on the bases of
the reversible equations of classical and quantum physics is raised.
I will not discuss the solutions of this problem in classical physics, 
I will examine the quantum mechanics. First of all  it is necessary to note that the question about irreversibility of quantum processes   
can be examined in connection with the problem of interpretation of quantum mechanics. Briefly, I will examine two most important  approaches - Copenhagen and  Everett's  interpretations. \\
Copenhagen interpretation of quantum mechanics asserts that we cannot~speak about the quantum properties of the system  before these properties are measured~\cite{been},~\cite{chaos}. In other words, quantum theory describes not quantum world but the fact that we can speak about quantum world after measurement. Let the energy of the quantum system has two values  $E_{1}$,  $E_{2}$ and  $\varphi_{1}$,  $\varphi_{2}$ corresponding wave functions. If the system is described by the wave function $\psi$ it is possible to present it in the form of the superposition of the functions $\varphi_{1}$ and  $\varphi_{2}$: $\psi =a_{1}\varphi_{1}+a_{2}\varphi_{2}$. Before measurement the wave function $\psi$ occupies simultaneously two levels and energy of the system does not have a specific value. Only after the measurement we will obtain the values $E_{1}$ or $E_{2}$ with probabilities 
$\mid a_{1}\mid^{2}$, $\mid a_{2}\mid^{2}$ respectively. The passage from the "potential possibilities"  which are described by the wave function $\psi$ to the "actual realities", which can be measured, is called reduction or collapse of the wave function. 
The physical process leading to collapse is called  quantum decoherence. The process of the measurement leads to the abrupt, irreversible change of the state. Any measurement is not a reversed  process.   
Thus, irreversibility appears when  the measurement  is carried out. In this case the behavior of the quantum system is described by Neumann's postulate of reduction~\cite{qm}  which, however, does not answer the question which {\it{dynamic processes}} lead to irreversibility.
Thus,  in accordance with Copenhagen interpretation of quantum mechanics the observed processes proceeding in the world are caused by our measurements including, for example, the quantum transitions. However, as Prigogine notes, since the quantum transitions are the basic mechanism of chemical reactions it is difficult to agree with the latter assertion. Is it possible  to consider the chemistry as the result of our observation? If yes, then who observed the chemical reactions which led to the appearance of life~\cite{ps}? \\
Another interpretation of quantum mechanics is proposed by H. Everett (relative state interpretation)~\cite{everet}.  It was developed in the works~\cite{witt},~\cite{deu} (many-world interpretation);~\cite{men0},~\cite{men},~\cite{men2} (extension of Everett's concept).  
Many-world interpretation allows the existence of the infinite numbers of classical realities: each term of quantum superposition  corresponds to one classical world. According to the extended  Everett's concept the measurement entangles the measured system with the environment, however, linear superposition is not destroyed. Thus, after measurement the linearity of quantum mechanics is preserved. 
In this case the reduction of the wave function does not occur. However, at the specific moment of time the observer sees only one result of the measurement~-~the one classical world. This occurs because the consciousness of the observer divides the state of quantum world into the classical realities which he receives independently. The consciousness of the observer subjectively makes the selection of one alternative. The consciousness continuously creates classical reality~\cite{men}. Hence follows: the classical world  is the illusion, reduction of the wave function is the illusion too, since they appear subjectively in the consciousness of the observer. In the approach, the concepts of the "arrow of time" and irreversibility are rather connected with its subjective feeling of the observer  then with the objective property of material itself. It is assumed, the further study of this interpretation  will be possible after the construction of the model of "quantum consciousness", which is discussed in the work~\cite{men2}.\\
Obviously, the approaches examined above contain a subjective, anthropomorphic element. The interpretations contain
a basic distinction between the quantum system and the observer. The latter fact is undesirable for the theory since it takes us away from the solution of the basic problem: the determination of the objective laws of the nature (if  we believe in the objectivity of the  nature). If Copenhagen interpretation is the direct consequence of the insufficient realization of the physical sense of the basic postulates of quantum mechanics, then  Everett's interpretation is a natural consequence of them. This situation impels one to the alternative formulation of quantum dynamics which does not appeal to the observer excluding the subjective, anthropomorphous element. In the approach  the irreversibility must be represented as the property of material itself and is not defined by the active role of the observer. \\
In the paper  I examine the alternative formulation of quantum dynamics in the framework of the Brussels-Austin group works that has been headed by I. Prigogine for many years. In the works, a possible variant of description of nonequilibrium processes at microscopic level in the frame of the Liouville space extension of quantum mechanics is investigated. The mechanism of the asymmetry of processes in the time which made it possible to accomplish a passage from the reversible evolution to  irreversible one was developed. Thus, new irreversible dynamics with the disrupted  time symmetry was formulated. The symmetry in the time disrupted as a result of asymmetric nature of the physically permissible states.\\
 At the present moment, it is
necessary to continue further development of the Brussels - Austin group approach in the framework of the realistic models of interaction. Then, the irreversible evolution of a charged spin $1/2$ particle interacting with a magnetic field is investigated on the basis of subdynamics theory.
In the work the survey of the basic principles of the Brussels-Austin group is carried out. 
In section 2 the eigenvalue problem is discussed. Perturbative solution of the Schroedinger equation in the framework of  complex spectral representation is given. In section 3  the Liouville formulation of quantum mechanics is represented. The task of the complex spectral representation of the Liouvillian is solved in section 4. In section 5 the theory of subdynamics is discussed.  The time evolution of the density matrix is determined. The interacting model of positive charged spin $1/2$ particle with magnetic field is examined in section 6. The complex eigenvalue problem for the spin system is solved in section 7. The expression for the density matrix element describing the irreversible evolution of the charged particle spin system and numerical calculations are obtained in section 8.

\section{Perturbative solution of the Schroedinger equation,  complex spectral representation}
As the first step, in the framework of Schroedinger equation, I examine as the asymmetry in time can arise. 
I examine the eigenvalue problem for the Hamiltonian
$H=H_{0}+\lambda V$
\begin{equation}\label{h22}
H\mid\psi_{\alpha}>=\tilde{E}_{\alpha}\mid\psi_{\alpha}>,
\end{equation}
where $H_{0}$ - free Hamiltonian, $\lambda V$ - interaction part, $\lambda$ - coupling constant. In the conventional case, Hamiltonian
$H$ is a Hermitian operator, $\tilde{E}_{\alpha}$ is a perturbed energy of the state - a real number. It is known that
the usual procedure of equation~\eqref{h22} solution on the basis of perturbation method can lead to the appearance of
the denominators $1/(E_{\alpha}-E_{\alpha'})$, where $E_{\alpha}$, $E_{\alpha'}$ are the energies corresponding
to the unperturbed situation. Obviously, the divergences can arise at $E_{\alpha}=E_{\alpha'}$.  The basic question now is - what we can do to avoid the divergences, when $E_{\alpha} =  E_{\alpha'}$.  
I examine the situation using the simple Friedrichs model~\cite{frid} (the model is presented closely to the text of the works~\cite{ppt}~-~\cite{pop3}). Despite the fact, that the solution of the problem for the Friedrichs model
is known it serves as a good example for the demonstration of the essence of situation. The model describes
interaction of two level atom and electromagnetic field. In the Friedrichs model $\mid1>$ corresponds to the atom in
its bare exited level~\cite{liuv1},~$\mid{k}>$ corresponds to the bare field mode with the atom in its ground state. The
state $\mid1>$ is coupled to the state $\mid{k}>$
\begin{equation}\label{f}
\begin{split}
&H=H_{0}+\lambda V \\
&=E_{1}\mid 1><1\mid+\sum_{k}E_{k}\mid k><k\mid +
\lambda\sum_{k}V_{k}(\mid k><1\mid + \mid 1><k\mid),
\end{split}
\end{equation}
where
\begin{equation}\label{f2}
\mid 1><1\mid + \sum_{k}\mid k><k\mid =1,~<\alpha\mid\alpha '>=\delta_{\alpha\alpha '}.
\end{equation}
Here $\alpha~(\alpha')=1$  or $k$. In the ordinary approach
the eigenvalue problem for the Hamiltonian $H$ is formulated as follows
\begin{equation}\label{h}
H\mid\psi_{1}>=\tilde{E_{1}}\mid\psi_{1}>.
\end{equation}
For the eigenstate $\mid\psi_{1}>$ (for small $\lambda $) perturbation
method gives the expression
\begin{equation}\label{f}
\mid\psi_{1}>\approx\mid 1>-\sum_{k}\frac{\lambda
V_{k}}{E_{k}-E_{1}}\mid k>.
\end{equation}
If $E_{1}>0$, divergences appear at $E_{k}=E_{1}$.
I. Prigogine and co-workers noted that in the general case the satisfactory solution of this
problem is impossible on the basis of the conventional formulation of quantum dynamics.  
However, the eigenvalue problem can be solved if the time ordering of the eigenstates will
be introduced. This procedure can be realized through the
introduction into the denominators imaginary terms:
$-i\varepsilon$ for the relaxation processes, which are oriented
into the future and $+i\varepsilon$ for the excitation processes,
which are oriented into the past. 
In this case the eigenvalue
problem~\eqref{h} is reduced to the complex eigenvalue problem
\begin{align}\label{ht1}
H
\mid\varphi_{1}>=Z_{1}\mid\varphi_{1}>,~<\widetilde{\varphi}_{1}\mid
H = <\widetilde{\varphi}_{1}\mid Z_{1},
\end{align}
where we must distinguish right - eigenstate $\mid\varphi_{1}>$ and left - eigenstate
$<\widetilde{\varphi}_{1}\mid$~\cite{ps,ppt,liuv1}. 
$Z_{1}$ is a complex
\begin{equation}\label{gamma}
Z_{1}={\bar{E}_{1}} - i\gamma.
\end{equation}
Here $\bar{E}_{1}$ is a renormalized energy and $\gamma$ is a real
positive value. This procedure makes it possible to avoid the divergences and leads to the following expressions
for the eigenstates $\mid\varphi_{1}>$, $<\widetilde{\varphi}_{1}\mid$~\cite{ppt}
\begin{equation}\label{f43}
\mid\varphi_{1}>\approx\mid 1>-\sum_{k}\frac{\lambda
V_{k}}{(E_{k}-\bar{E}_{1}- z)^{+}_{-i\gamma}}\mid k>,
\end{equation}
\begin{equation}\label{f42}
<\widetilde{\varphi}_{1}\mid\approx<1\mid-\sum_{k}\frac{\lambda
V_{k}}{(E_{k}-\bar{E}_{1}- z)^{+}_{-i\gamma}}<k\mid.
\end{equation}
In the
expressions~\eqref{f43} and~\eqref{f42} the designation
$1/(E_{k}-\bar{E}_{1}- z)^{+}_{-i\gamma}$ has been referred to as
"delayed analytic continuation"~\cite{ppt,liuv1} and can be determined 
through the integration with a test function $f(E_{k})$. For example,
\begin{equation}\label{ht300}
\begin{split}
\int\limits_{0}^{\infty}dE_{k}\frac{f(E_{k})} {(E_{k}-\bar{E}_{1}-
z)^{+}_{-i\gamma}}\equiv \lim\limits_{z\rightarrow-i\gamma}
\Bigr{(}\int\limits_{0}^{\infty}dE_{k}\frac{f(E_{k})} {E_{k}-
\bar{E}_{1}~-~z}\Bigl{)}_{z\in C^{+}},
\end{split}
\end{equation}
where we first have to evaluate the integration on the upper
half-plane $C^{+}$ and then the limit of $z\rightarrow-i\gamma$
must be taken. Thus, the approach makes it possible to solve the problems which could not be solved within the framework of classical and quantum mechanics. Now we can realize the program of Heisenberg - to solve task at the eigenvalues which could not be solved within the framework of traditional methods. 
The spectral representation of the
Hamiltonian takes the form
\begin{equation}\label{hqw}
H=\sum_{\alpha}Z_{\alpha}\mid\varphi_{\alpha}><{\tilde{\varphi}_{\alpha}}\mid.
\end{equation}
For the eigenstates we have relations
\begin{equation}\label{ht}
\sum_{\alpha}\mid\varphi_{\alpha}><{\tilde{\varphi}_{\alpha}}\mid
= 1,~
<{\tilde{\varphi}_{\alpha}}\mid\varphi_{\alpha'}>=\delta_{\alpha\alpha'}.
\end{equation}
Since, $H$ is Hermitian the
corresponding eigenstates $\mid\varphi_{1}>$, $<\widetilde{\varphi}_{1}\mid$ are outside Hilbert space, have no
Hilbert norm
\begin{align}\label{htrr}
&<\varphi_{1}\mid\varphi_{1}>=<{{\tilde{\varphi}}_{1}}\mid{\tilde{\varphi}_{1}}>=0
\end{align}
and furthermore have a zero average energy
\begin{align}\label{htrr2}
<\varphi_{1}\mid H\mid\varphi_{1}>=<{{\tilde{\varphi}}_{1}}\mid H\mid{\tilde{\varphi}_{1}}>=0.
\end{align}
The eigenstates $\mid\varphi_{1}>$, $<\widetilde{\varphi}_{1}\mid $ are called "Gamow
vectors". Extensive literature is dedicated to the study of "Gamow
vectors", for example~\cite{gam1}~-~\cite{gam4}.\\ 
Thus, the Hermiticity of $H$ leads to the fact that "usual" norms of eigenstates
$\mid\varphi_{1}>$, $<\tilde{\varphi}_{1}\mid$ disappear. However, the eigenstates $\mid\varphi_{1}>$,
$<{\tilde{\varphi}_{1}}\mid$ have a broken time symmetry. We can associate $\mid\varphi_{1}>$ with the unstable state,
which vanishes for $t\rightarrow +\infty$, $\mid{\tilde{\varphi}_{1}}>$ corresponds to the state, which vanishes for
$t\rightarrow - \infty$
\begin{equation}\label{hq}
\mid\varphi_{1}(t)>=\exp(-i{\bar{E}_{1}}t-\gamma
t)\mid\varphi_{1}(0)>,
\end{equation}
\begin{equation}\label{hq}
\mid\tilde{\varphi}_{1}(t)>=\exp(-i{\bar{E}_{1}}t+\gamma
t)\mid\tilde{\varphi}_{1}(0)>.
\end{equation}
However, the are few unpleasant aspects. \\
Using the eigenvectors $\mid\varphi_{1}>$, $\mid{\tilde{\varphi}_{1}}>$ it is possible to construct the density operators~\cite{liuv1}
\begin{equation}\label{oper}
\rho_{a}=\mid{\tilde{\varphi}_{1}}><\varphi_{1}\mid ,~\rho_{b}=\mid\varphi_{1}><{{\tilde{\varphi}}_{1}}\mid .
\end{equation}
They are invariants of motion and they do not describe the process of relaxation. In addition, "unusual" norms and zero average energy ~\eqref{htrr},~\eqref{htrr2} exist.\\
These are the reasons why a more general space of density matrix-Liouville space will be examined.
Nevertheless, it is important to emphasize: the complex perturbative solution of the eigenvalue problem in the framework of the Schroedinger equation makes it possible to determine $\gamma$ which has important physical sense. The value $2\gamma$ determines the probability of transition between two quantum states~\cite{liuv1},~\cite{shirm}. This result will be important for us in the next sections.

\section{Liouville formalism}
Now I will examine the equation which determines the density matrix $\rho$ - Liouville-von Neumann equation 
\begin{equation}\label{ter}
i\frac{\partial\rho}{\partial t}= L\rho.
\end{equation}
Liouville-von Neumann operator (Liouvillian) has the form:
\begin{align}\label{liuvil}
L=H\times 1 - 1\times H,
\end{align}
here symbol "$\times$" denotes the operation $(A\times
B)\rho$=A$\rho$B. Operator $L$
is written down in the sum of free part $L_{0}$, that depends
on the free Hamiltonian $H_{0}$ and the interaction part $L_{I}$,
that depends on $\lambda V\equiv H_{I}$ - Hamiltonian of interaction: $L=L_{0} + L_{I}$.
Let state
$|\alpha>$ be a eigenvector of the free Hamiltonian, then  we have the equation
$H_{0}|\alpha>$=$\omega_{\alpha}|\alpha>$ with the
energy $\omega_{\alpha}$. The dyad of the states
$|\alpha ><\beta|$ is the eigenstate of operator $L_{0}$

\begin{equation}\label{liouv}
L_{0}|\alpha><\beta|=(\omega_{\alpha}-\omega_{\beta})|\alpha ><\beta|
\end{equation}
or 
\begin{equation}\label{liouv2}
L_{0}|\alpha ;\beta\rangle\rangle=w_{\alpha,\beta}|\alpha;
\beta\rangle\rangle,
\end{equation}
where the designations
$|\alpha >< \beta|\equiv |\alpha; \beta\rangle\rangle$
and $w_{\alpha,\beta}\equiv\omega_{\alpha}-\omega_{\beta}$ were used. \\
In the Liouville
space for the dyadic operators we have the relations:\\
the inner product defined by
\begin{align}\label{j}
\langle\langle A|B\rangle\rangle \equiv Tr(A^{\dag}B),
\end{align}
the matrix elements are given by
\begin{align}\label{j2}
< \alpha|A|\beta> \equiv \langle\langle \alpha;
\beta|A\rangle\rangle,
\end{align}
the biorthogonality and bicompleteness relations have the form:
\begin{align}\label{j3}
\langle\langle \alpha^{'};\beta^{'}|\alpha;\beta\rangle\rangle =
\delta_{\alpha' \alpha}\delta_{\beta'
\beta},~\sum\limits_{\alpha,\beta}|\alpha;\beta\rangle\rangle\langle\langle\alpha;\beta|=1.
\end{align} 
For  operator $L$ we have the equation (the text is written close to
the materials of the works~\cite{ps,liuv1})
\begin{equation}\label{lv2}
L\mid f_{\alpha,\beta}\rangle\rangle =\tilde{w}_{\alpha,\beta}\mid f_{\alpha,\beta}\rangle\rangle,
\end{equation}
here  $\mid f_{\alpha,\beta}\rangle\rangle\equiv\mid\psi_{\alpha}><\psi_{\beta}\mid $, 
$\tilde{w}_{\alpha,\beta}=\tilde{E}_{\alpha}-\tilde{E}_{\beta}$ - the eigenvector and eigenvalue of Liouville-von Neumann operator $L$, where $\mid\psi_{\alpha}>$ 
($\mid\psi_{\beta}>$),  $\tilde{E}_{\alpha}$ ($\tilde{E}_{\beta}$)
- the eigenvector and eigenvalue of the total Hamiltonian~\eqref{h22}.
The spectral representation of operator $L$ has the form
\begin{equation}\label{rep}
L=\sum\limits_{\alpha ,\beta}\mid f_{\alpha,\beta}\rangle\rangle\tilde{w}_{\alpha,\beta}\langle\langle f_{\alpha,\beta}\mid.
\end{equation}
Since $\tilde{w}_{\alpha,\alpha}=\tilde{E}_{\alpha}-\tilde{E}_{\alpha}=0$, the diagonal elements of density matrix do not depend on the time. Therefore, strictly speaking, the conventional formalism does not make it possible to describe the evolution of the system to the equilibrium. 
Eq.~\eqref{lv2} can be rewritten by using the correlation index $\nu$:
\begin{equation}\label{lv}
L\mid f_{\nu}\rangle\rangle=\tilde{w}^{\nu}\mid f_{\nu}\rangle\rangle,
\end{equation}
where $\tilde{w}^{\nu}\equiv \tilde{w}_{\alpha,\beta}$ ( or ${w}^{\nu}\equiv {w}_{\alpha,\beta}$ for the case~\eqref{liouv2})  and $\mid f_{\nu}\rangle\rangle\equiv \mid f_{\alpha,\beta}\rangle\rangle$.
The details of the theory of correlations can be found, for example, in the works~\cite{ps,pp3,corel}.
In equation~\eqref{lv}  $\nu = 0$ if $\alpha=\beta$ -
that is the set of diagonal operators $\mid\alpha><\alpha\mid$ - "vacuum of correlations" and $\nu\neq 0$ is the remaining off-diagonal case. Then, the  concept of degree of correlation must be introduced. The degree of correlation of dyads $\mid\nu\rangle\rangle$, $d$, has been defined as the minimum number of interactions $L_{I}$  by which a dyadic state can reach the vacuum of correlation. For example, the degree of vacuum of correlation $d=0$.  
The concept of correlation will be examined  lately in connection with the model which is investigated in the article. \\
Certainly, the problem of the description of irreversibility can be solved on the basis of well known physical approximations~\cite{blum}. The solution is based on the distinction between the open system and its environment. The environment is assumed to be in the thermodynamic equilibrium. However, as it was noted~\cite{dec}
"This distinction introduces an anthropomorphic element. Indeed, the environment, introduced by one observer, may be different to that seen by another observer. Moreover, this distinction is phenomenological as one avoids answering the most fundamental question of  nonequilibrium statistical mechanics, namely, what is the mechanism that governs the approach to thermal equilibrium of the environment, without violating the basic laws of physics."  
The approach of Brussels-Austin group is based on the assertion: the irreversibility is the objective property of
quantum world. This objective reality we have to describe without including the anthropomorphic principle, such as lack of our knowledge, coarse-graining approximations, environmental approach~\cite{decog}.
The solution of the problems, which was examined above, is proposed on the basis of the {\it Liouville space extension} 
of quantum mechanics~\cite{liuv1},~\cite{liuv2} or otherwise, on the basis of {\it the complex spectral representation} of Liouville-von Neumann operator~$L$, and {\it subdynamics  theory}.  In this case, Liouvillian has complex eigenvalues, that break time symmetry without
introduction of an anthropomorphic element or
dissipative terms to the Hamiltonian~\cite{kad} or to the Liouvillian~\cite{Isar}.

\section{Complex spectral representation of Liouvillian}
We have a new formulation of the eigenvalue problem: 
\begin{align}\label{ff3u}
L\mid\Psi^{\nu}_{j}\rangle\rangle=Z^{\nu}_{j}\mid\Psi^{\nu}_{j}\rangle\rangle,~ \langle\langle
\tilde{\Psi}^{\nu}_{j}\mid L=\langle\langle \tilde{\Psi}^{\nu}_{j}\mid Z^{\nu}_{j},
\end{align}
where $Z^{\nu}_{j}$ are the complex values, $\nu$ is a correlation index and $j$ is a degeneracy index, since one type of correlation index can correspond to the different states
 (the complex eigenvalue problem for the Liouvillian is examined in the
works~\cite{liuv1,pp9,pp10}). 
It was shown~\cite{liuv1,corel,crit2} that the eigenstates of $L$ can be written in the
terms of kinetic operators $C^{\nu}$ and $D^{\nu}$. Operator
$C^{\nu}$ creates correlations other than the $\nu$ correlations,
$D^{\nu}$ is destruction operator. The use
of the kinetic operators allows to write down the expressions for the
eigenstates of Liouville operator in the following form~\cite{liuv1}
\begin{align}\label{ae6}
|\Psi^{\nu}_{j}\rangle\rangle=(N^{\nu}_{j})^{1/2}\Phi^{\nu}_{C}|u^{\nu}_{j}\rangle\rangle
,~\langle\langle\widetilde{\Psi}^{\nu}_{j}|=\langle\langle
\widetilde{v}^{\nu}_{j}|\Phi^{\nu}_{D}(N^{\nu}_{j})^{1/2},
\end{align}
where
\begin{align}\label{ae7}
\Phi^{\nu}_{C}\equiv P^{\nu} + C^{\nu},~\Phi^{\nu}_{D}\equiv
P^{\nu} + D^{\nu},
\end{align}
here $N^{\nu}_{j}$ - is a normalization constant. The
determination of the states $|u^{\nu}_{j}\rangle\rangle$,
$\langle\langle\widetilde{v}^{\nu}_{j}|$ and operators $P^{\nu}$,
$C^{\nu}$, $D^{\nu}$ can be found in the works~\cite{liuv1,liuv2,pp3}. In
the general case, for example, the operators $P^{\nu}$ are
determined by the following expressions~\cite{pp3}
\begin{align}\label{ae84}
P^{\nu}=\sum\limits_{j}|u^{\nu}_{j}\rangle\rangle \langle\langle\widetilde{u}^{\nu}_{j}|,
~\langle\langle\widetilde{u}^{\nu}_{j}|u^{\nu'}_{j'}\rangle\rangle=\delta_{\nu\nu'}\delta_{j j'}.
\end{align}
Let me examine some results obtained in the original works.
The work~\cite{liuv1} shows that $\Phi^{\nu}_{C}$, $\Phi^{\nu}_{D}$
satisfy so-called nonlinear Lippmann-Schwinger equation. For the
$\Phi^{\nu}_{C}$ we have
\begin{align}\label{fieq}
\Phi^{\nu}_{C}=P^{\nu}+\sum\limits_{\mu\neq\nu}
P^{\mu}\frac{-1}{w^{\mu}-w^{\nu}-i\varepsilon_{\mu\nu}}[L_{I}\Phi^{\nu}_{C}-\Phi^{\nu}_{C}L_{I}\Phi^{\nu}_{C}]P^{\nu},
\end{align}
where the time ordering $-i\varepsilon_{\mu\nu}$ was introduced.
This is the important moment of the theory because the time ordering breaks time symmetry and, thus, leads to the description of the irreversibility. 
The determination of the sign of the infinitesimals $\varepsilon_{\mu\nu}$ is connected 
with degree of correlation $d$. In accordance with our experience we will consider the directions to the
higher degrees of correlation oriented in the future, for example, the decay of elementary particles, quantum transitions with emission or absorption of  energy, and the
directions to the lowest degrees of correlation are oriented in
the past. This leads to the relations:
\begin{align}\label{tetad}
\varepsilon_{\mu\nu}=+\varepsilon~ \text{if}~d_{\mu}\geq
d_{\nu}~(t>0);~\varepsilon_{\mu\nu}=-\varepsilon~
\text{if}~d_{\mu}< d_{\nu}~(t<0).
\end{align}
For the $\Phi^{\nu}_{D}$ we have the equation
\begin{align}\label{fieq2}
\Phi^{\nu}_{D}=P^{\nu}+P^{\nu}[\Phi^{\nu}_{D}L_{I}-\Phi^{\nu}_{D}L_{I}\Phi^{\nu}_{D}]
\sum\limits_{\mu\neq\nu}
P^{\mu}\frac{1}{w^{\nu}-w^{\mu}-i\varepsilon_{\nu\mu}}.
\end{align}
Eqs.~\eqref{fieq},~\eqref{fieq2} determine the kinetic operators
of creation $C^{\nu}$ and destruction $D^{\nu}$ as follows
\begin{equation}\label{cdeqq}
C^{\nu}=\sum\limits_{\mu\neq\nu}
P^{\mu}\frac{-1}{w^{\mu}-w^{\nu}-i\varepsilon_{\mu\nu}}[L_{I}\Phi^{\nu}_{C}-\Phi^{\nu}_{C}L_{I}\Phi^{\nu}_{C}]P^{\nu},
\end{equation}
\begin{equation}\label{kjhg}
D^{\nu}=P^{\nu}[\Phi^{\nu}_{D}L_{I}-\Phi^{\nu}_{D}L_{I}\Phi^{\nu}_{D}]
\sum\limits_{\mu\neq\nu}
P^{\mu}\frac{1}{w^{\nu}-w^{\mu}-i\varepsilon_{\nu\mu}}.
\end{equation}
It is evident from the last expressions that
$D^{\nu}\neq(C^{\nu})^{\dag}$, but $D^{\nu}=(C^{\nu})^{\ast}$, where "$\dag$" corresponds to the Hermitian conjugation
and the action "$\ast$"~-~the "star" conjugation
which is Hermitian conjugation plus the change
$\varepsilon_{\mu\nu}\rightarrow \varepsilon_{\nu\mu}$.\\
Substituting the expression~\eqref{ae6} in Eq.~\eqref{ff3u} and multiplying $P^{\nu}$ from left on both sides, we
obtain~\cite{liuv1}
\begin{align}\label{ae8}
\theta^{\nu}_{C}|u^{\nu}_{j}\rangle\rangle =
Z^{\nu}_{j}|u^{\nu}_{j}\rangle\rangle,
\end{align}
where
\begin{align}\label{ae9}
\begin{split}
&\theta^{\nu}_{C}\equiv P^{\nu}L(P^{\nu} + C^{\nu})=L_{0}P^{\nu} +
P^{\nu}L_{I}(P^{\nu} + C^{\nu})P^{\nu}=\\
&w^{\nu}P^{\nu} +
P^{\nu}L_{I}\Phi^{\nu}_{C}P^{\nu}.
\end{split}
\end{align}
Also, it is possible to show the validity of the following useful relationship 
\begin{equation}\label{lf}
L \Phi^{\nu}_{C}=\Phi^{\nu}_{C}\theta^{\nu}_{C}.
\end{equation}
The latter will be necessary for us in the future.
In Eq.~\eqref{ae8} $\theta^{\nu}_{C}$ is the collision operator connected with the kinetic operator $C^{\nu}$. This is
non-Hermitian dissipative operator which plays the main role in the nonequilibrium dynamics. As it was shown in the
work~\cite{pp3} operator $\theta^{0}_{C}$ can be reduced to the collision operator in Pauli master equation for the
weakly coupled systems. Comparing Eqs.~\eqref{ff3u},~\eqref{ae8} we can see that $|u^{\nu}_{j}\rangle\rangle$ is
eigenstate of collision operators $\theta^{\nu}_{C}$ with the same eigenvalues $Z^{\nu}_{j}$ as $L$. \\
Thus, the time ordering  introduces into the theory the "arrow of the time", and leads 
to the connection of quantum mechanics with kinetic, time irreversible dynamics.
As an example, I examine the Friedrichs model.
In  the works ~\cite{liuv1},~\cite{pop3} the following results were obtained: 
\begin{equation}\label{ty}
\begin{split}
&L\mid\Psi^{0}_{1}\rangle\rangle=Z^{0}_{1}\mid\Psi^{0}_{1}\rangle\rangle,~~(\nu=0), \\
&Z^{0}_{1}=-2i\gamma,~~\mid\Psi^{0}_{1}\rangle\rangle =\mid\varphi_{1};\varphi_{1}\rangle\rangle;
\end{split}
\end{equation}
\begin{equation}\label{ty2}
\begin{split}
&L\mid\Psi^{\nu}\rangle\rangle=Z^{\nu}\mid\Psi^{\nu}\rangle\rangle,~~( \nu\neq 0),\\
&Z^{1,k}=Z_{1}-E_{k},~~Z^{k,1}=E_{k}-Z^{c.c.}_{1},\\
&Z^{k,k'}=E_{k}-E_{k'},
\end{split}
\end{equation}
where, for example,
\begin{equation}\label{ty4}
\begin{split}
\mid\Psi^{k,k'}\rangle\rangle=\mid{\varphi}_{k};{\varphi}_{k'}\rangle\rangle.\\
\end{split}
\end{equation}
In~\eqref{ty}-\eqref{ty4} $c.c.$ is the operation of the complex conjugation, $Z_{1}$  is determined by relationship~\eqref{gamma},
 $\mid\varphi_{1}>$ is determined by~\eqref{f43} and $\mid{\varphi}_{k}>$, $\mid\tilde{\varphi}_{k}>$ can be found in the works ~\cite{liuv1},~\cite{pop3} and correspond to the field mode.\\
Obviously, the complex eigenvalues $Z^{\nu}_{j}$ of the Liouville operator cannot be reduced to the simple difference of two eigenvalues of the Hamiltonian as in the case~\eqref{lv2}. Furthermore, eigenstates $\mid\Psi^{\nu}_{j}\rangle\rangle$, $\langle\langle \tilde{\Psi}^{\nu}_{j}\mid$  are not the multiplications of the wave functions, they cannot be obtained from the Schroedinger equation. In this sense we have the {\it alternative formulation of quantum theory}.

\section{The theory of subdynamics, time evolution of the density matrix}
Equation~\eqref{ter} does not describe the irreversible evolution. The precise description of irreversibility requires the determination of the new approach. This approach was developed on the basis of the theory of subdynamics. 
In the theory of Brussels-Austin group, subdynamics is called the
construction of a complete set of spectral projectors $\Pi^{\nu}$~\cite{ppt,liuv1,pop3} (see also~\cite{ph,g3,hgm7})
\begin{align}\label{poper}
\Pi^{\nu}=\sum\limits_{j}|\Psi^{\nu}_{j}\rangle\rangle\langle\langle\widetilde{\Psi}^{\nu}_{j}|.
\end{align}
The projectors $\Pi^{\nu}$ satisfy the following relations
\begin{equation}\label{com}
\begin{split}
&\Pi^{\nu}L=L\Pi^{\nu},~(\text{commutativity});~
\sum\limits_{\nu}\Pi^{\nu}=1,~(\text{completeness});\\
&\Pi^{\nu}\Pi^{\nu'}=\Pi^{\nu}\delta_{\nu\nu'},~(\text{orthogonality});
~\Pi^{\nu}=(\Pi^{\nu})^{\ast},~(\text{star-Hermiticity})
\end{split}
\end{equation}
and in our approximation 
\begin{align}\label{poper}
\Pi^{\nu}\rightarrow P^{\nu}~\text{if interaction is absent ($\lambda$=0)}.
\end{align}
Operator $\Pi^{\nu}$ can be represented in the following
form
\begin{align}\label{pca3}
\Pi^{\nu}=(P^{\nu}+C^{\nu})A^{\nu}(P^{\nu}+D^{\nu}),
\end{align}
where $A^{\nu}$ is the star-Hermitian operator $A^{\nu}=(A^{\nu})^{\ast}$
\begin{align}\label{pca2}
A^{\nu}=P^{\nu}(P^{\nu}+D^{\nu}C^{\nu})^{-1}P^{\nu}.
\end{align}
Taking~\eqref{com} it is possible to write down the density matrix
$\rho$ in the form
\begin{align}\label{poper}
\rho(t)=\sum\limits_{\nu}\Pi^{\nu}\rho(t)=\sum\limits_{\nu}\rho^{\nu}(t),
\end{align}
where $\rho^{\nu}\equiv\Pi^{\nu}\rho$. 
Using~\eqref{ter},~\eqref{lf},~\eqref{com}  and~\eqref{pca3} it is possible to write down 
\begin{equation}\label{e11}
\begin{split}
&\rho^{\nu}(t)=\Pi^{\nu}\exp(-iLt)\rho(0)=\\
&(P^{\nu}+C^{\nu})\exp(-i\vartheta^{\nu}_{C}t)A^{\nu}(P^{\nu}+D^{\nu})\rho(0).
\end{split}
\end{equation}
I examine the component $P^{\nu}\rho^{\nu}(t)$. It was  called the "privileged" component of $\rho^{\nu}(t)$.
For the $P^{\nu}\rho^{\nu}(t)$ it is easy to obtain 
\begin{equation}\label{e100}
\begin{split}
P^{\nu}\rho^{\nu}(t)=P^{\nu}\exp(-i\vartheta^{\nu}_{C}t)A^{\nu}(P^{\nu}+D^{\nu})\rho(0),
\end{split}
\end{equation}
where in our case $\rho(0)=P^{0}\rho(0)$ - vacuum of correlations and the relations $P^{\nu}C^{\nu}=0$, $P^{\nu}P^{\nu}=P^{\nu}$  were  used.
The relation~\eqref{e100} leads to the equation
\begin{equation}\label{e110}
\begin{split}
i\frac{\partial P^{\nu}\rho^{\nu}(t)}{\partial t}=\vartheta^{\nu}_{C}P^{\nu}\rho^{\nu}(t).
\end{split}
\end{equation}
Thus, it is shown operators $\rho^{\nu}$ satisfy
separate equations of motion. The equation~\eqref{e110} is the kinetic equation for each $\Pi^{\nu}$ subspace.
This is the reason why the
projectors $\Pi^{\nu}$ can be associated with the introduction of
the concept of subdynamics.\\
Using~\eqref{poper}  I determine the component $P^{0}\rho(t)$ 
\begin{equation}\label{uuu}
P^{0}\rho(t)=P^{0}\sum\limits_{\nu}\rho^{\nu}(t)=P^{0} \rho^{0}(t)+ P^{0} \sum\limits_{\nu\neq 0}\rho^{\nu}(t).
\end{equation}
The expression $P^{0} \rho^{0}(t)$ is determined by~\eqref{e100} for $\nu=0$. The second term has the form
\begin{equation}\label{e7}
\begin{split}
&P^{0}\sum\limits_{\nu\neq 0}\rho^{\nu}(t)=\sum\limits_{\nu\neq 0}P^{0}(P^{\nu}+C^{\nu})\exp(-i\vartheta^{\nu}_{C}t)A^{\nu}(P^{\nu}+D^{\nu})\times\\
&P^{0}\rho(0)=\sum\limits_{\nu\neq 0}P^{0}C^{\nu}\exp(-i\vartheta^{\nu}_{C}t)A^{\nu}D^{\nu}P^{0}\rho(0).
\end{split}
\end{equation}
The obtained expressions lead 
\begin{equation}\label{pro}
\begin{split}
P^{0}\rho(t)=P^{0}\exp(-i\vartheta^{0}_{C}t)A^{0}P^{0}\rho(0)+
\sum\limits_{\nu\neq 0}P^{0}C^{\nu}\exp(-i\vartheta^{\nu}_{C}t)A^{\nu}D^{\nu}
P^{0}\rho(0).
\end{split}
\end{equation}
Taking the time derivative of  $P^{0}\rho(t)$ - component  we have
\begin{equation}\label{e150}
\begin{split}
&i\frac{\partial P^{0}\rho(t)}{\partial t}=\vartheta^{0}_{C}P^{0}\rho^{0}(t)+\\
&\sum\limits_{\nu\neq 0}P^{0}C^{\nu}\vartheta^{\nu}_{C}\exp(-i\vartheta^{\nu}_{C}t)A^{\nu}D^{\nu}P^{0}\rho(0).
\end{split}
\end{equation}
The equation~\eqref{e150} consists of two parts: the first one corresponds to the Markovian approximation, the second part includes the memory effects and determines the non-Markovian  processes~\cite{dec}. \\
It is possible to conclude, the introduction of the subdynamics makes it possible to determine the Eqs.~\eqref{e110} and~\eqref{e150} which are the fundamental equations for the description of the irreversible processes. 

\section{ Charged particle spin behavior in the magnetic field} 
I examine the behavior of the positive charged particle in the magnetic field~\cite{abragam}. As the charged particle it is possible to examine proton, positron and any positive charged particle with non-zero magnetic moment and with spin $1/2$. \\
Spin  $\vec{S}$ of the particle is connected with magnetic moment $\vec{\mu}$ by the expression
\begin{equation}\label{e1}
\begin{split}
\vec{\mu}=\gamma_{p}\vec{S},
\end{split}
\end{equation} 
where the values $\gamma_{p}$, $\vec{S}$ are determined by
\begin{equation}\label{e11}
\begin{split}
\gamma_{p} = g_{p}\frac{e}{2m_{p}},~~\vec{S}=\frac{1}{2}\vec{\sigma}.
\end{split}
\end{equation} 
Here $e$,  $m_{p}$ are the charge and  the mass of  particle, $g_{p}$ is $g$-factor, $\vec{\sigma}$ -
Pauli matrices. I use the system of units, where $\hbar =1$ and the speed of light $c=1$.\\
Let the isolated charged particle with magnetic moment $\vec{\mu}$
be placed into the external magnetostatic field  which is directed along  $z$ axis - $H_{z}$.
The Hamiltonian operator then becomes 
\begin{equation}\label{e2}
\begin{split}
{{H}}_{SH_{z}}=-g_{p}\frac{e}{4m_{p}}{\sigma}_{z}H_{z}.
\end{split}
\end{equation}
In the expression~\eqref{e2} ${\sigma}_{z}$ is a $z$-component of the vector $\vec{\sigma}$.\\
Particle with spin $1/2$ in the external magnetic  field $H_{z}$ has two energy levels corresponding to two different values of magnetic quantum number  $m$
\begin{equation}\label{e456}
\begin{split}
E_{m}=-m\gamma_{p}H_{z},~\text{where}~ m=\pm 1/2.
\end{split}
\end{equation}
The distance between the levels is determined by the energy $\Delta E$ 
\begin{equation}\label{e6}
\begin{split}
\Delta E =\omega_{0}=\gamma_{p}H_{z}.
\end{split}
\end{equation}
The transitions of the particle from one energy level to another $E_{-1/2}\leftrightarrows E_{1/2}$  can occur as a result of the influence on the system of the external magnetic field $\vec{H}_{1}$  perpendicular to the field $H_{z}$. The exciting magnetic field  $\vec{H}_{1}$ must be variable and its frequency $\omega$ must coincide with frequency $\omega_{0}$. Such transitions are accompanied by emission or absorption of energy $\Delta E$. This is the well known phenomenon of the magnetic resonance.\\
Strictly speaking, the energy levels are not determined accurately. They have an uncertainty of the values. It leads to the spectrum of the radiated energies, which distribution is simulated by the introduction of distribution function $f(\omega)$ with the width $\Delta$. The  distribution function describes Lorentzian or Gaussian distribution of energy. Lorentzian distribution is described by the function~\cite{abragam}
\begin{equation}\label{e867}
\begin{split}
f(\omega)=\frac{\delta}{\pi}\frac{1}{\delta^{2}+(\omega -\omega_{0})^{2}},~\delta=\frac{1}{2}\Delta
\end{split}
\end{equation}
with norm
\begin{equation}\label{norm}
\int\limits_{-\infty}^{\infty}f(\omega)d\omega = 1.
\end{equation}
Taking into account the distribution $f(\omega)$ it is possible to determine the rate of the transition between the energy levels ~\cite{abragam},~\cite{emsli}
\begin{equation}\label{f88}
\begin{split}
W_{1/2\rightarrow -1/2}=W_{-1/2\rightarrow 1/2}\equiv W=\frac{1}{2}\pi\gamma_{p}^{2}H_{1}^{2}f(\omega).
\end{split}
\end{equation}
In the simplest case, the components of the field $\vec{H}_{1}$ change according to the law~\cite{slonim}
\begin{equation}\label{e22} 
H_{x}=H_{1}\cos(\omega t),~~H_{y}=-H_{1}\sin(\omega t),
\end{equation}
where $\omega$ is the frequency of the rotating field in the $xy$ plane.  
Then, the Hamiltonian operator takes the form 
\begin{equation}\label{e1}
\begin{split}
&{H}_{magnetic}=-g_{p}\frac{e}{4m_{p}}\vec{\sigma}\cdot\vec{H} =\\
&-g_{p}\frac{e}{4m_{p}}\left(\sigma_{x} H_{x} + \sigma_{y} H_{y} + \sigma_{z}H_{z}\right).
\end{split}
\end{equation}
Whence it follows 
\begin{equation}\label{e3}
{H}_{magnetic}=-g_{p}\frac{e}{4m_{p}}\begin{pmatrix}H_{z} & H_{x}-iH_{y} \\ H_{x}+iH_{y} & -H_{z}\end{pmatrix}.
\end{equation}
Since,
\begin{equation}\label{e44}
H_{x}-iH_{y}=H_{1}\exp(i\omega t),~~H_{x}+iH_{y}=H_{1}\exp(-i\omega t) 
\end{equation}
I obtain the expression 
\begin{equation}\label{e55}
\begin{split}
&{H}_{magnetic}=-g_{p}\frac{e}{4m_{p}}\sigma_{z}H_{z} - g_{p}\frac{e}{4m_{p}}\begin{pmatrix}0& H_{1}\exp(i\omega t)  \\ H_{1}\exp(-i\omega t) & 0\end{pmatrix}\equiv \\
& {H}_{SH_{z}} +{H}_{SH_{xy}}.
\end{split}
\end{equation}
The first term ${H}_{SH_{z}}$ in the expression~\eqref{e55} determines interaction with the field $H_{z}$, the second term ${H}_{SH_{xy}}$ determines interaction with the field $\vec{H}_{1}$.\\
I examine the field $H_{1}$ so that $H_{1}\ll H_{z}$ (see  A. Abragam~\cite{abragam}). This condition has the greatest practical interest. 
In this case for the eigenfunctions $\phi_{m}$, which are determined by the equation 
\begin{equation}\label{poi}
{H}_{magnetic}\phi_{m} =\tilde{E}_{m}\phi_{m},~~m=\pm 1/2,
\end{equation} 
I take the approximation
\begin{equation}\label{ert}
\phi_{1/2} = \begin{pmatrix}1 \\ 0\end{pmatrix}\exp(-iE_{1/2} t),~~\phi_{-1/2}=\begin{pmatrix}0 \\ 1\end{pmatrix}\exp(-iE_{-1/2} t).
\end{equation}
Then, the function, which describes the behavior of spin, can be determined by the superposition of the functions $\phi_{1/2}$, $\phi_{-1/2}$ 
\begin{equation}\label{psi}
\begin{split}
&\phi = a_{1/2}\phi_{1/2} + a_{-1/2}\phi_{-1/2}=\\
&a_{1/2}\begin{pmatrix}1 \\ 0\end{pmatrix}\exp(-iE_{1/2} t) + a_{-1/2}\begin{pmatrix}0 \\ 1\end{pmatrix}\exp(-iE_{-1/2} t).
\end{split}
\end{equation}
For the Hermitian conjugated ($h.c.$) function $\varphi^{h.c.}$ I have 
\begin{equation}\label{e18}
\begin{split}
&\phi^{h.c.} = a^{h.c.}_{1/2}\phi^{h.c.}_{1/2} + a^{h.c.}_{-1/2}\phi^{h.c.}_{-1/2} = \\
&a_{1/2}^{h.c.}\begin{pmatrix}1 & 0 \end{pmatrix}\exp(iE_{1/2} t) + a_{-1/2}^{h.c.}\begin{pmatrix}0 & 1 \end{pmatrix}\exp(iE_{-1/2} t).
\end{split}
\end{equation}
The expression~\eqref{e55} corresponds to the monochromatic case. 
For the passage to the general case it is necessary to make the following replacements 
\begin{equation}\label{e7}
\begin{split}
&H_{1}\exp(i\omega t) \rightarrow \int\limits_{-\infty}^{\infty}A(\omega)H_{1}\exp(i\omega t)d\omega ,\\
&H_{1}\exp(-i\omega t) \rightarrow \int\limits_{-\infty}^{\infty}B(\omega)H_{1}\exp(-i\omega t)d\omega ,
\end{split}
\end{equation}
where the value of the coefficients $A(\omega)$, $B(\omega)$ will be determined later.

\section{Complex eigenvalue problem for the spin system.}
I examine the transition $E_{-1/2}\rightarrow E_{1/2}$.  This transition  
is accompanied by the emission of the energy $\Delta E$~\eqref{e6}. The transition with the emission of the energy is the irreversible process.
The description of the irreversible evolution of the spin system will be carried out on the basis of the expression~\eqref{pro}.
I will determine the evolution of matrix element $<-1/2\mid\rho (t)\mid -1/2>$ 
which corresponds to the probability of finding the particle in the state with spin $-1/2$ at the time moment $t$.\\
The evolution of the system in the kinetic - Markovian approximation has the exponential nature $\backsim\exp(-w t)$. So, in the work~\cite{shirm} (in this work the irreversible evolution of the unstable $\pi^{-}$ - meson was defined)  $w\equiv 2\gamma_{\mathbf{p}_{\pi}}$ - the rate of the  $\pi^{-}$ - meson decay. I assume that in our case $w\equiv 2\gamma_{-1/2}=W$, where $W$ is the rate~\eqref{f88} under the resonance condition $\omega = \omega_{0}$ ($\omega_{0}$ is determined by~\eqref{e6}).
Therefore, the first step will consist of the determination of the value $\gamma_{-1/2}$. This task will be executed on the basis of perturbative
solution of complex eigenvalue problem in the framework of the method of the second quantization.\\
Let me examine the replacement~\eqref{e7}.
According to the method of the second quantization I define the values $A(\omega)$, $B(\omega)$ as operators  
\begin{equation}\label{e667}
\begin{split}
A(\omega)\rightarrow g(\omega)a^{\dagger}(\omega),~~B(\omega)\rightarrow g(\omega)a(\omega).
\end{split}
\end{equation}
Here $a^{\dagger}(\omega)$, $a(\omega)$ are the operators of creation and destruction of photon with the energy $\omega$.
I assume, these operators satisfy the commutation relation
\begin{equation}\label{e8}
\begin{split}
[a(\omega),a^{\dagger}(\omega ')]=\delta(\omega - \omega ') .
\end{split}
\end{equation}
In the determination~\eqref{e667}, $g(\omega)$ is the function satisfying the condition 
\begin{equation}\label{e9}
\int\limits_{-\infty}^{\infty}g^{2}(\omega)d\omega =1.
\end{equation}
As will be shown below (see~\eqref{change}) the square of the function $g(\omega)$  is the distribution function~\eqref{e867}.
The coefficients $a_{\pm1/2}$ in the expression~\eqref{psi}  find the value of the destruction operators of particles with  $m=\pm1/2$. For $a_{\pm1/2}^{h.c.}$ and $\phi^{h.c.}$  in~\eqref{e18} we must make
$a_{\pm1/2}^{h.c.}\rightarrow a_{\pm1/2}^{\dagger}$, $\phi^{h.c.}\rightarrow \phi^{\dagger}$, here $a_{\pm1/2}^{\dagger}$ are the creation operators. It is assumed, the operators  $a_{\pm1/2}$,
$a_{\pm1/2}^{\dagger}$ satisfy the anti-commutation relation
\begin{equation}\label{e212}
\begin{split}
[a_{m},a_{m'}^{\dagger}]_{+} = \delta_{mm'},~~m~(m')=\pm 1/2.
\end{split}
\end{equation}
In this case the Hamiltonian of the particle-magnetic field system can be determined in the form
\begin{equation}\label{e665}
\begin{split}
&H_{magnetic}= -g_{p}\frac{e}{4m_{p}}\phi^{\dagger}\sigma_{z}H_{z}\phi  -\\
&g_{p}\frac{e}{4m_{p}}\phi^{\dagger}\begin{pmatrix}0& \int\limits_{-\infty}^{\infty}g(\omega)a^{\dagger}(\omega)H_{1}\exp(i\omega t)d\omega  \\  \int\limits_{-\infty}^{\infty}g(\omega)a(\omega) H_{1}\exp(-i\omega t)d\omega & 0\end{pmatrix}\phi \\=
&{H}_{SH_{z}} +{H}_{SH_{xy}},
\end{split}
\end{equation}
where the previous designations of operators are preserved.
Now, the values $\phi$, $\phi^{\dagger}$ and Hamiltonian~\eqref{e665}  can be examined as the operators in the Dirac representation.\\
I examine the eigenvalue problem for the Hamiltonian
$H=H_{0}+H_{I}$, where
a free part $H_{0}=H_{\omega}+{H}_{SH_{z}}$, and an interaction part $H_{I}\equiv\lambda V={H}_{SH_{xy}}$. $H_{\omega}$ is a free electromagnetic Hamiltonian. Let state $\mid\omega>=a^{\dagger}(\omega)\mid 0>$ ($\mid 0>$ is the vacuum state: $a(\omega)\mid 0>=0$) be a eigenvector of the free Hamiltonian $H_{\omega}$ then, $H_{\omega}\mid\omega>=\omega\mid\omega>$, where $\omega$ is the photon energy. The Hamiltonian $H_{\omega}$
can be determined as follows 
\begin{equation}\label{e665po}
\begin{split}
H_{\omega}=\int\limits_{-\infty}^{\infty}\omega a^{\dagger}(\omega)a(\omega)d\omega.
\end{split}
\end{equation}
As will be shown below (appendix C), for our model, here and in the subsequent expressions
integration over  $\omega$ have to be carried out in the interval $\omega \in [0,~\infty)$.
 I determine the states  $\mid\pm 1/2> =a_{\pm1/2}^{\dagger}\mid 0>$. The states $\mid\pm 1/2>$ are the  eigenvectors of the  Hamiltonian ${H}_{SH_{z}}$ 
\begin{equation}\label{e665ki}
\begin{split}
{H}_{SH_{z}}= -g_{p}\frac{e}{4m_{p}}\phi^{\dagger}\sigma_{z}H_{z}\phi=\sum\limits_{m=1/2,-1/2}E_{m}a_{m}^{\dagger}a_{m}
\end{split}
\end{equation}
and ${H}_{SH_{z}}\mid\pm 1/2>=E_{\pm 1/2}\mid\pm 1/2>$. In this case the expression  $H_{0}\mid \pm1/2, \omega>=(\omega + E_{\pm 1/2})\mid \pm1/2, \omega>$ is valid, where  $\mid \pm1/2, \omega>=a_{\pm1/2}^{\dagger}a^{\dagger}(\omega)\mid 0>$.
Since, the transition ($E_{-1/2}\rightarrow E_{1/2}$) occurs as the result of interaction with the field $\vec{H}_{1}$ that 
is why the term ${H}_{SH_{xy}}$ is defined as the interaction part $\lambda V$.\\
I will solve
the problem assuming, that the eigenvalue $Z_{-1/2}$ of the
Hamiltonian $H$ is complex (the general formalism of the complex
spectral representation can be found in the works~\cite{ppt,liuv1,pop3,dfg,ap3}). 
In accordance
with the approach~\cite{ppt}, in our case, I will distinguish
equation for the right eigenstate $\mid\varphi_{-1/2}>$ and
for the left eigenstate $<\widetilde{\varphi}_{-1/2}\mid$ of
Hamiltonian $H$
\begin{align}\label{ht}
H\mid\varphi_{-1/2}>=Z_{-1/2}\mid\varphi_{-1/2}>,
~<\widetilde{\varphi}_{-1/2}\mid H = <\widetilde{\varphi}_{-1/2}\mid
Z_{-1/2}.
\end{align}
I expand the values $\mid\varphi_{-1/2}>$,
$<\widetilde{\varphi}_{-1/2}\mid$, $Z_{-1/2}$ in the
perturbation series
\begin{align}\label{ht3}
&\mid\varphi_{-1/2}>=\sum\limits_{n=0}^{\infty}\lambda^{n}\mid\varphi_{-1/2}^{(n)}>,~
<\widetilde{\varphi}_{-1/2}\mid=\sum\limits_{n=0}^{\infty}\lambda^{n}<\widetilde{\varphi}_{-1/2}^{(n)}\mid, \\
&Z_{-1/2}=\sum\limits_{n=0}^{\infty}\lambda^{n}Z_{-1/2}^{(n)},
\end{align}
where
\begin{align}\label{ht4}
\begin{split}
&\mid\varphi_{-1/2}^{(0)}>=\mid -1/2>,~
<\widetilde{\varphi}_{-1/2}^{(0)}\mid=< -1/2 \mid,~
Z_{-1/2}^{(0)}=E_{-1/2},\\
&~\lambda \equiv e-\text{ the charge of particle.}
\end{split}
\end{align}
In accordance with the definitions~\eqref{ht}-\eqref{ht4} for the first equation of~\eqref{ht} I can write down 
\begin{equation}\label{e11}
\begin{split}
&<-1/2\mid\Bigl{(}H_{0}\sum\limits_{n=0}^{\infty}e^{n}\mid\varphi_{-1/2}^{(n)}> + e V\sum\limits_{n=0}^{\infty}e^{n}\mid\varphi_{-1/2}^{(n)}>\Bigr{)} =\\
&<-1/2\mid\sum\limits_{n=0}^{\infty}e^{n}Z_{-1/2}^{(n)}\sum\limits_{n'=0}^{\infty}e^{n'}\mid\varphi_{-1/2}^{(n')}>,
\end{split}
\end{equation}
where $\mid -1/2>$ is our initial state.\\
Hence, it is easy to obtain the expression for the coefficient $Z_{-1/2}^{(n)}$ 
\begin{equation}\label{e15}
\begin{split}
Z_{-1/2}^{(n)} =<-1/2\mid V\mid\varphi_{-1/2}^{(n-1)}> - \sum\limits_{l=1}^{n-1}Z_{-1/2}^{(l)}<-1/2\mid\varphi_{-1/2}^{(n-l)}>.
\end{split}
\end{equation}
Let us examine the first term in the expression~\eqref{e15}. The use of the interaction term in~\eqref{e665} leads to 
\begin{equation}\label{e250}
\begin{split}
&<-1/2\mid V\mid\varphi_{-1/2}^{(n-1)}> = -\frac{g_{p}H_{1}}{4m_{p}}\int\limits_{-\infty}^{\infty} g(\omega)\exp{(-i(E_{1/2}-E_{-1/2}+\omega)t)}\times \\
&<1/2,\omega\mid\varphi_{-1/2}^{(n-1)}>d\omega .
\end{split}
\end{equation}
The substitution of the expression~\eqref{e250} into~\eqref{e15} and summing up $\sum\limits_{n=1}^{\infty}e^{n}$ on the both sides of the obtained expression lead to the result 
\begin{equation}\label{rte}
\begin{split}
&\sum\limits_{n=1}^{\infty}e^{n}Z_{-1/2}^{(n)} = -\frac{g_{p}H_{1}}{4 m_{p}}\int\limits_{-\infty}^{\infty}  g(\omega)
\sum\limits_{n=1}^{\infty}e^{n}\exp{(-i(E_{1/2}-E_{-1/2}+\omega)t)}\times\\
&<1/2,\omega\mid\varphi_{-1/2}^{(n-1)}>d\omega - \sum\limits_{n=1}^{\infty}e^{n}\sum\limits_{l=1}^{n-1}Z_{-1/2}^{(l)}<-1/2\mid\varphi_{-1/2}^{(n-l)}>
\end{split}
\end{equation}
whence I obtain 
\begin{equation}\label{e26}
\begin{split}
&Z_{-1/2} - E_{-1/2} =  -e\frac{g_{p}H_{1}}{4 m_{p}}\int\limits_{-\infty}^{\infty}  g(\omega)\exp{(-i(E_{1/2}-E_{-1/2}+\omega)t)}\times \\
&<1/2,\omega\mid\varphi_{-1/2}>d\omega -
(Z_{-1/2} - E_{-1/2})(<-1/2\mid\varphi_{-1/2}> -1).
\end{split}
\end{equation}
Let us determine the value $<1/2,\omega\mid\varphi_{-1/2}>$.
For this purpose I write down the obvious equality
\begin{equation}\label{e5}
\begin{split}
&<1/2, \omega\mid\Bigr{(}H_{0}\sum\limits_{n=0}^{\infty}e^{n}\mid\varphi_{-1/2}^{(n)}> + e V \sum\limits_{n=0}^{\infty}e^{n}\mid\varphi_{-1/2}^{(n)}>\Bigr{)} = \\
&<1/2, \omega\mid\sum\limits_{n=0}^{\infty}e^{n}Z_{-1/2}^{(n)}\sum\limits_{n'=0}^{\infty}e^{n'}\mid\varphi_{-1/2}^{(n')}>,
\end{split}
\end{equation}
where $\mid 1/2, \omega>$ is our final state. \\
The expression~\eqref{e5} leads to
\begin{equation}\label{e10}
\begin{split}
&<1/2,\omega\mid\varphi_{-1/2}^{(n)}> =  \frac{-1}{E_{1/2}+\omega - E_{-1/2}}\times\\
&\Bigl{(}<1/2,\omega\mid V\mid\varphi_{-1/2}^{(n-1)}> - \sum\limits_{l=1}^{n}Z_{-1/2}^{(l)}<1/2,\omega\mid\varphi_{-1/2}^{(n-l)}> \Bigr{)}.
\end{split}
\end{equation}
It is assumed, the state with the spin $-1/2$ disappears in the future, converting to the state with the spin $1/2$ and photon with the energy $\omega$. In this case the time ordering of the expression~\eqref{e10} has to be introduced. For this purpose I make 
the replacement 
\begin{equation}\label{e103}
\begin{split}
\frac{1}{E_{1/2}+\omega - E_{-1/2}}\rightarrow\frac{1}{E_{1/2}+\omega - E_{-1/2}-i\varepsilon_{\mu\nu}}.
\end{split}
\end{equation}
Here the symbol $\mu$ in infinitesimal $\varepsilon_{\mu\nu}$ corresponds to the state $\mid  1/2,\omega>$, the symbol $\nu$ corresponds to the state $\mid -1/2>$.
The state $\mid  1/2,\omega>$ has the high degree of correlation and I assume $\varepsilon_{\mu\nu}\equiv\varepsilon >0$.\\
For the first term of~\eqref{e10} I have the result  
\begin{equation}\label{e200}
\begin{split}
&<1/2,\omega\mid V\mid\varphi_{-1/2}^{(n-1)}> = -\frac{g_{p}H_{1}}{4m_{p}}g(\omega)
\exp(i(E_{1/2}-E_{-1/2}+\omega)t) \times\\
&<-1/2\mid\varphi_{-1/2}^{(n-1)}>.
\end{split}
\end{equation}
The substitution of the expressions~\eqref{e200},~\eqref{e103} into~\eqref{e10} and summing up $\sum\limits_{n=1}^{\infty}e^{n}$ lead to 
\begin{equation}\label{e223}
\begin{split}
&<1/2,\omega\mid\varphi_{-1/2}> = e\frac{g_{p}H_{1}}{4 m_{p}}g(\omega) \exp(i(E_{1/2}-E_{-1/2}+\omega)t) \times\\
&<-1/2\mid\varphi_{-1/2}>\frac{1}{E_{1/2}+\omega - Z_{-1/2} - i\varepsilon} .
\end{split}
\end{equation}
Taking into account the result~\eqref{e223}, the expression~\eqref{e26} can be represented in the form 
\begin{equation}\label{e27}
\begin{split}
Z_{-1/2} = E_{-1/2} - \left( \frac{eg_{p}H_{1}}{4m_{p}}\right) ^{2}\int\limits_{-\infty}^{\infty} g^{2}(\omega)\frac{1}{E_{1/2}+\omega - Z_{-1/2} -i\varepsilon} d\omega .
\end{split}
\end{equation}
Using the formal expression
\begin{equation}\label{e28}
\begin{split}
\frac{1}{E_{1/2}+\omega - Z_{-1/2} -i\varepsilon}  \approx \textbf{P}\frac{1}{E_{1/2}+\omega -E_{-1/2} } + i\pi\delta (E_{1/2}+\omega - E_{-1/2}) 
\end{split}
\end{equation}
I present the eigenvalue $Z_{-1/2}$ as follows 
\begin{equation}\label{e293}
\begin{split}
Z_{-1/2} = \bar{E}_{-1/2} -i\gamma_{-1/2},
\end{split}
\end{equation}
where $ \bar{E}_{-1/2}$ is a renormalized energy
\begin{equation}\label{e30}
\begin{split}
&\bar{E}_{-1/2} = E_{-1/2} - \left( \frac{eg_{p}H_{1}}{4m_{p}}\right) ^{2}~\textbf{P}\int\limits_{-\infty}^{\infty} g^{2}(\omega)\frac{1}{E_{1/2}+\omega - E_{-1/2}} d\omega, \\
&\text{\textbf{P} - the principal part  of the integral}
\end{split}
\end{equation}
and
\begin{equation}\label{e31}
\begin{split}
\gamma_{-1/2} = \frac{1}{4}\pi\gamma_{p}^{2}H_{1}^{2}g^{2}(\omega_{0}) .
\end{split}
\end{equation}
Comparing the obtained expression~\eqref{e31} for $\gamma_{-1/2}$ with the expression~\eqref{f88}  for the rate $W$ we see 
\begin{equation}\label{e34}
\begin{split}
W_{-1/2\rightarrow 1/2} /_{\omega = \omega_{0}} =2 \gamma_{-1/2}/_{g^{2}(\omega_{0})= f(\omega_{0})} = \frac{1}{2}\pi\gamma_{p}^{2}H_{1}^{2}f(\omega_{0}),
\end{split}
\end{equation}
where the equality
\begin{equation}\label{change}
g^{2}(\omega)= f(\omega)
\end{equation}
is used. 
Evidently, in this case, the obtained expression $2\gamma_{-1/2}$ coincides with the expression for $W_{-1/2\rightarrow 1/2}$ under the resonance condition $\omega_{0}=E_{-1/2}-E_{1/2}$. 

 \section{Charged particle spin evolution }
Let us  examine the matrix element $\langle\langle -1/2;-1/2\mid P^{0}\rho(t) \rangle \rangle$ 
(see the expression ~\eqref{pro}).
The projection operator $P^{0}$~\eqref{ae84} can be redetermined in the form~\eqref{e17}.
For the matrix element the following designations are valid
\begin{equation}\label{e188}
\begin{split}
&\langle \langle -1/2 ; -1/2\mid P^{(0)}\rho(t)\rangle \rangle = \langle \langle -1/2; -1/2\mid \rho(t)\rangle \rangle \equiv  \\
&< -1/2\mid \rho (t)\mid -1/2>.
\end{split}
\end{equation}
 For the operator $A^{\nu}$~\eqref{pca2} the equality
\begin{equation}\label{e20}
\begin{split}
A^{\nu}=P^{\nu}(P^{\nu}+D^{\nu}C^{\nu})^{-1}P^{\nu} = P^{\nu} + \sum\limits_{n=1}^{\infty}(-1)^{n}(D^{\nu}C^{\nu})^{n}
\end{split}
\end{equation} 
is correct.
Then, in the first approximation, it is possible to assume that 
\begin{equation}\label{e21}
A^{\nu}\approx P^{\nu}.
\end{equation}
Using the expressions~\eqref{pro},~\eqref{e188} and~\eqref{e21} I  obtain 
\begin{equation}\label{e19}
\begin{split}
&\langle \langle -1/2;-1/2\mid\rho(t)\rangle \rangle =\langle \langle -1/2;-1/2\mid\exp(-i\vartheta^{0}_{C}t)P^{0}\rho(0)\rangle \rangle +\\
&\sum\limits_{\nu\neq 0}\langle \langle -1/2; -1/2\mid C^{\nu}\exp(-i\vartheta^{\nu}_{C}t)P^{\nu}D^{\nu}P^{0}\rho(0)\rangle \rangle. \end{split}
\end{equation}
First of all, I examine the first term in the expression~\eqref{e19}. This term corresponds to the Markovian approximation. 
In our case the initial condition is given by
\begin{equation}\label{incon}
\rho(0)=\mid -1/2><-1/2\mid,~~~~ \text{i.e.,}~ \langle \langle -1/2;-1/2\mid\rho(0)\rangle \rangle =1,
\end{equation}
and for our matrix element I have 
\begin{equation}\label{yui}
\begin{split}
\langle \langle -1/2;-1/2\mid &\exp(-i\vartheta^{0}_{C}t)P^{0}\rho(0)\rangle \rangle = \\
&\langle \langle -1/2;-1/2\mid\exp(-i\vartheta^{0}_{C}t)\mid-1/2;-1/2\rangle \rangle.
\end{split}
\end{equation}
From the expression~\eqref{ae9} it follows
\begin{equation}\label{e25}
\vartheta^{0}_{C} = P^{0}L_{I}C^{0}P^{0},
\end{equation}
here it was assumed that
\begin{equation}\label{e24}
 P^{0}L_{I} P^{0}=0.
\end{equation}
According to~\cite{ph}  operator $C^{\nu}$ (the expression~\eqref{cdeqq}), in the first approximation can be represented in the form 
\begin{equation}\label{e32}
\begin{split}
C^{\nu}=\sum\limits_{\mu\neq\nu}P^{\mu}\frac{-1}{w^{\mu}-w^{\nu}-i\varepsilon_{\mu\nu}}L_{I}P^{\nu}.
\end{split}
\end{equation}
Hence, for the operator $C^{0}$ I obtain 
\begin{equation}\label{e325}
\begin{split}
C^{0}=\sum\limits_{\mu\neq 0}P^{\mu}\frac{-1}{w^{\mu}-i\varepsilon_{\mu 0}}L_{I}P^{0}.
\end{split}
\end{equation}
Here  the symbol $\nu =0$ corresponds  to the vacuum of correlations, $\mu$ in infinitesimal $\varepsilon_{\mu 0}$ corresponds to the  high degree of correlation.
Infinitesimal determines the time ordered transition from the vacuum of correlations to the high degree of correlation.
In our model the initial condition~\eqref{incon} cuts from the possible vacuum of correlation states the initial correlation state 
$\mid -1/2;-1/2\rangle \rangle$ which corresponds to initial pure state $\mid -1/2>$.
High degree of correlations $\mu$ are determined by the correlation final state $\mid 1/2~ \omega ; -1/2\rangle \rangle$ which corresponds to the final state $\mid 1/2,\omega>$. In this case  $\varepsilon_{\mu 0}\equiv\varepsilon >0$.
The details of the evaluation of the expression for the matrix element~\eqref{yui} are examined in the appendix A:
\begin{equation}\label{mark}
\begin{split}
\langle \langle -1/2;-1/2\mid \exp(-i\vartheta^{0}_{C}t)P^{0}\rho(0)\rangle \rangle = \exp(-2\gamma_{-1/2}t).
\end{split}
\end{equation}
The non-Markovian effect is determined by the second term in the expression~\eqref{e19}.
For it I have the result
\begin{equation}\label{e2967}
\begin{split}
&\sum\limits_{\nu\neq 0}\langle \langle -1/2; -1/2\mid C^{\nu}\exp(-i\vartheta^{\nu}_{C}t)P^{\nu}D^{\nu}P^{0}\rho(0)\rangle \rangle =\\
&{\Bigr{(}\frac{eg_{p}H_{1}}{4m_{p}}\Bigl{)}}^{2}\Bigr{(}\int\limits_{-\infty}^{\infty}f(\omega)\frac{\exp(-i(\omega -\omega_{0})t)}
{(\omega - \omega_{0} - i\varepsilon)^{2}}d\omega + c.c.\Bigl{)}.
\end{split}
\end{equation}
The details of calculations of the expression~\eqref{e2967} are represented in the appendix B.\\
Taking into account the determination ~\eqref{e188} and the results~\eqref{mark},~\eqref{e2967} the time evolution of the charged particle spin system can be represented  by the expression 
\begin{equation}\label{e476}
\begin{split}
&< -1/2\mid \rho (t)\mid -1/2> =\\
&\exp(-2\gamma_{-1/2}t) + 
{\Bigr{(}\frac{eg_{p}H_{1}}{4m_{p}}\Bigl{)}}^{2}\Bigr{(}\int\limits_{-\infty}^{\infty}f(\omega)\frac{\exp(-i(\omega -\omega_{0})t)}
{(\omega - \omega_{0} - i\varepsilon)^{2}}d\omega + c.c.\Bigl{)}. 
\end{split}
\end{equation}
The calculation of the integral in the expression~\eqref{e476} (the features of the calculation are examined in the appendix C)  leads to the final result.
\begin{equation}\label{e4769}
\begin{split}
&< -1/2\mid \rho (t)\mid -1/2>= \\
&\exp(-2\gamma_{-1/2}t) + 
\frac{\gamma_{p}^{2}H_{1}^{2}}{2\delta\pi \omega_{0}^{2}t}\Bigl{(} A(t)\sin(\omega_{0}t) + B(t)\cos(\omega_{0}t)\Bigr{)}.
\end{split}
\end{equation}
The coefficients $A(t)$, $B(t)$ in~\eqref{e4769} are determined by the expressions
\begin{equation}\label{e4152}
\begin{split}
A(t) = \int\limits_{0}^{\infty}d\xi\exp(-\xi)
\frac{a^{4}(a^{4}(1+b) - a^{2}(1+6b)\xi^{2}+b\xi^{4})}{D(\xi)},
\end{split}
\end{equation}
\begin{equation}\label{e51648}
\begin{split}
B(t) =  \int\limits_{0}^{\infty}d\xi\exp(-\xi)
\frac{-a^{7}(2+4b)\xi + 4a^{5}b\xi^{3}}{D(\xi)}.
\end{split}
\end{equation}
The function $D(\xi)$ has the form
\begin{equation}\label{e41525}
\begin{split}
&D(\xi)=a^{8}(1+b)^{2} +a^{6}(2+b(2+4b))\xi^{2}+a^{4}(1+b(-2+6b))\xi^{4}+\\
&a^{2}b(-2+4b)\xi^{6}+b^{2}\xi^{8},
\end{split}
\end{equation}
where $a=\omega_{0}t$ and $b=(\omega_{0}/\delta)^{2}$.\\
Thus, the privileged component in the determination~\eqref{uuu} corresponds to the kinetic, Markovian process (the first term in~\eqref{e4769}), the non-Markovian effect coming from the nonprivileged component (the second term in~\eqref{e4769}). \\
Figure 1 presents the calculations of the separate contributions: the Markovian  and the non-Markovian terms. Calculations are executed for proton: proton mass - $938.27~MeV$,~$g$-factor - 5.58,~nuclear magneton - $ 3.15\times 10^{-14}~MeV~T^{-1}$ \cite{emsli},~\cite{pph}.
It is seen, the relaxation process is determined by the  Markovian term. The kinetic evolution  prevails in the entire time interval. The non-Markovian term is determined by the damped oscillations and  is not essential. The other sets  of the values ($H_{z}$, $H_{1}$, $\delta$) substantially decrease the contribution of the latter.
Note, that the calculation of  the non-Markovian effect was carried out earlier, for example, in the work~\cite{decog} (a quantum harmonic oscillator linearly coupled to a bosonic massless scalar field), where the contribution of the latter is more  noticeable (see also~\cite{dec},~\cite{decoger2}). A time scale of transition of the evolution from the non-Markovian regime to the Markovian regime was determined. Thus, it was shown, the non-Markovian term corresponds to the well-known quantum Zeno effect~\cite{sud}.

\section{Conclusion}
In conclusion it is necessary to note that the approach I have examined is not the one more additional interpretation of quantum mechanics. It is not intended to interpret the quantum mechanics, but it is proposed to reformulate it so that the irreversibility naturally would enter into quantum dynamics. The approach denies the conventional opinion that the irreversibility appears only at the macroscopic level, while the microscopic level can be described by the laws, reversed in the time. Thus, the irreversible nature of quantum physics is asserted. In this case, the basic problem, designated by Boltzmann and Planck - to formulate the second law of thermodynamics at the microscopic level, is revealed. In the approach the solution of this problem is proposed in the works~\cite{been},~\cite{pp10},~\cite{hf1}~-~\cite{hf4}. 
The law of the increase of entropy is accepted as fundamental there, that determines the
"arrow of time", the difference between the past and the future. The future corresponds to the larger value of the entropy. Thus, 
 the {\it existence of the "arrow of time", the second law of thermodynamics at the microscopic level are determined as the basic postulates}.
Unconditionally, the interaction with the environment leads to the irreversible effects, but in the Brussels-Austin group approach the environment is not limited by the special requirements as, for example, to be in the thermodynamic equilibrium. The preceding calculation of the relaxation time (Fig.1, Markovian term) for a spin $1/2$ particle interacting with a magnetic field demonstrates 
the utility of the Brussels-Austin formalism in a practical problem that may be tested in experiment.
\\
Is it possible to speak about the laws of quantum world independently of the observer? Bohr came to the conclusion about the impossibility of description of the independent quantum processes. The approach of  the Brussels-Austin group is the attempt to learn to describe the processes in the quantum world as the property of the material itself without active role of the observer. 

{\bf Acknowledgements}\\
I am grateful to Dr. A. S.~Kuchma, Dr. D. V.~Shulga  for the helpful
discussion and Dr. A. A. Goy, Dr. A. V.~Molochkov~for the
support of this work.

\setcounter{equation}{0}
\def\theequation{A.\arabic{equation}}
{\bf Appendix A. Markovian approximation }\\
In our approximation the projection operator $P^{0}$~\eqref{ae84} ($\nu=0$ - vacuum of correlations) can be redetermined in the form
\begin{equation}\label{e17}
\begin{split}
&P^{0}=\mid -1/2;-1/2\rangle \rangle\langle \langle -1/2;-1/2\mid +\\
&\frac{1}{\Omega}\int\limits_{-\infty}^{\infty}\mid 1/2 ~\omega ; 1/2~ \omega\rangle \rangle\langle \langle 1/2 ~\omega ; 1/2 ~\omega\mid d\omega,
\end{split}
\end{equation}
where the volume $\Omega$ is determined so that $\frac{\delta(0)}{\Omega}=1$ and $\delta(0)$ is the delta-function.
For the  operators $P^{\nu}$ ($\nu\neq 0$ - high degree of correlations ) I determine
\begin{equation}\label{e1779}
\begin{split}
&P^{\nu}=\int\limits_{-\infty}^{\infty}\mid -1/2;1/2\omega\rangle \rangle\langle \langle -1/2;1/2\omega\mid d\omega + \\
&\int\limits_{-\infty}^{\infty}\mid 1/2\omega;-1/2\rangle \rangle\langle \langle 1/2\omega;-1/2\mid d\omega.
\end{split}
\end{equation}
Then, the following relationships are valid:
\begin{equation}\label{e1796}
\begin{split}
P^{0}P^{0}...P^{0} = P^{0},~~~P^{\nu}P^{\nu}...P^{\nu} = P^{\nu},~~~P^{0}P^{\nu} = P^{\nu}P^{0}=0.
\end{split}
\end{equation}
The expansion of the expression~\eqref{yui} in the series leads to
\begin{equation}\label{e29}
\begin{split}
&\langle \langle -1/2;-1/2\mid\exp(-i\vartheta^{0}_{C}t)\mid -1/2;-1/2\rangle\rangle = \\
&1+(-it)\langle \langle -1/2;-1/2\mid \vartheta^{0}_{C}\mid -1/2;-1/2\rangle \rangle + \\
&\frac{(-it)^{2}}{2!}\langle \langle -1/2;-1/2\mid {(\vartheta^{0}_{C}})^{2}\mid -1/2;-1/2\rangle \rangle + \\
&\frac{(-it)^{3}}{3!}\langle \langle -1/2;-1/2\mid {(\vartheta^{0}_{C}})^{3}\mid -1/2;-1/2\rangle \rangle + ...~. 
\end{split}
\end{equation}
Taking into account ~\eqref{e25}  for the term $\langle \langle -1/2;-1/2\mid \vartheta^{0}_{C}\mid -1/2;-1/2\rangle \rangle$ I obtain
\begin{equation}\label{e1}
\begin{split}
&\langle \langle -1/2;-1/2\mid \vartheta^{0}_{C}\mid -1/2;-1/2\rangle \rangle =\\
&\langle \langle -1/2;-1/2\mid P^{0}L_{I}C^{0}P^{0}\mid -1/2;-1/2\rangle \rangle =\\
&\langle \langle -1/2;-1/2\mid L_{I}\sum\limits_{\mu\neq 0}P^{\mu}\frac{-1}{w^{\mu}-i\varepsilon_{\mu 0}}
L_{I} \mid -1/2;-1/2\rangle \rangle =\\
&\langle \langle -1/2;-1/2\mid L_{I}\sum\limits_{\alpha,\beta(\alpha\neq \beta)}P^{\alpha,\beta}\frac{-1}{w_{\alpha,\beta}-i\varepsilon}L_{I} \mid -1/2;-1/2\rangle \rangle,
\end{split}
\end{equation}
where $\varepsilon_{\mu 0}\equiv\varepsilon>0$. In our case
\begin{equation}\label{e296}
\begin{split}
&\sum\limits_{\alpha,\beta(\alpha\neq \beta)}P^{\alpha,\beta}\frac{-1}{w_{\alpha,\beta}-i\varepsilon}\equiv\\
&\int\limits_{-\infty}^{\infty}\mid 1/2 \omega ;-1/2\rangle \rangle\langle \langle 1/2\omega; -1/2\mid\frac{-1}{w_{1/2\omega, -1/2}-i\varepsilon}d\omega + \\
&\int\limits_{-\infty}^{\infty}\mid -1/2  ;1/2\omega\rangle \rangle\langle \langle -1/2; 1/2\omega\mid\frac{-1}{w_{-1/2,1/2\omega }-i\varepsilon}d\omega.
\end{split}
\end{equation}
Here,   $w_{1/2\omega,-1/2}=-w_{-1/2,1/2\omega} =E_{1/2}+\omega-E_{-1/2}$.
Using the determinations 
\begin{equation}\label{e234}
L_{I}=H_{I}\times 1 -1\times H_{I},~~(A\times B)C=ACB;
\end{equation}
\begin{equation}\label{e345}
\begin{split}
\langle \langle\alpha ;\beta \mid L_{I}\mid\alpha ' ;\beta '\rangle \rangle = <\alpha\mid H_{I}\mid\alpha '>\delta_{\beta '\beta } - \delta_{\alpha\alpha '}<\beta '\mid H_{I}\mid\beta>,
\end{split}
\end{equation}
where $\delta_{\beta '\beta }$ ($\delta_{\alpha\alpha '}$) is delta-function if the indices $\beta '$, $\beta$ ($\alpha$, $\alpha '$)
correspond to the continuous spectrum and it is Kronecker's symbol if the indices correspond to the discrete spectrum, for ~\eqref{e1} we will have
\begin{equation}\label{iuo}
\begin{split}
&\langle \langle -1/2;-1/2\mid \vartheta^{0}_{C}\mid -1/2;-1/2\rangle \rangle =\\
&-\int\limits_{-\infty}^{\infty} <-1/2\mid H_{I}\mid 1/2, \omega ><1/2, \omega\mid H_{I}\mid -1/2>\Bigl{(}\frac{1}{w_{1/2\omega,-1/2} - i\varepsilon} -\\
&\frac{1}{w_{1/2\omega, -1/2} + i\varepsilon}\Bigr{)}d\omega.
\end{split}
\end{equation}
Keeping in mind the result~\eqref{e293} it is possible to determine the equality 
\begin{equation}\label{jnm}
\begin{split}
&\frac{1}{E_{1/2}+\omega-Z_{-1/2}-i\varepsilon}\equiv\frac{1}{(E_{1/2}+\omega-z)^{+}_{-i\gamma_{-1/2}}}=\\
&\frac{1}{E_{1/2}+\omega-E_{-1/2}-i\varepsilon}+O(e),
\end{split}
\end{equation}
where $O(e)$ is the term of higher order on $e$.\\
In the equality~\eqref{jnm} the concept of "delayed analytic continuation" is used (section 2).
From the relationship~\eqref{jnm} another equality follows
\begin{equation}\label{jnm2}
\begin{split}
&\frac{1}{E_{1/2}+\omega-Z_{-1/2}^{c.c.}+i\varepsilon}\equiv\frac{1}{(E_{1/2}+\omega - z)^{-}_{+i\gamma_{-1/2}}}=\\
&\frac{1}{E_{1/2}+\omega-E_{-1/2}+i\varepsilon}+O(e).
\end{split}
\end{equation}
In this case the operation of integration of the expressions, which contain the values of the form $\frac{1}{(E_{1/2}+\omega - z)^{-}_{+i\gamma_{-1/2}}}$ is determined by the rule: we first have to evaluate the integration on the lower half-plane  and then the limit $z\rightarrow +i\gamma_{-1/2}$ must be taken. 
Taking into account~\eqref{jnm},~\eqref{jnm2}, using the Hamiltonian~\eqref{e665} and limiting  by order $e^{2}$ it is easy to obtain
\begin{equation}\label{e7779}
\begin{split}
&\langle \langle -1/2;-1/2\mid \vartheta^{0}_{C}\mid -1/2;-1/2\rangle \rangle = \\
&-\Bigl{(}\frac{eg_{p}H_{1}}{4m_{p}}\Bigr{)}^{2}\int\limits_{-\infty}^{\infty}g^{2}(\omega)\frac{1}{(E_{1/2}+\omega -z)^{+}_{-i\gamma_{-1/2}}}d\omega+\\
&\Bigl{(}\frac{eg_{p}H_{1}}{4m_{p}}\Bigr{)}^{2}\int\limits_{-\infty}^{\infty}g^{2}(\omega)\frac{1}{(E_{1/2}+\omega -z)^{-}_{+i\gamma_{-1/2}}}d\omega.
\end{split}
\end{equation}
Comparing the obtained expression~\eqref{e7779} with the expression~\eqref{e27} and  taking into account the designations~\eqref{e293},~\eqref{jnm},~\eqref{jnm2} I can write down 
\begin{equation}\label{e8779}
\begin{split}
&\langle \langle -1/2;-1/2\mid \vartheta^{0}_{C}\mid -1/2;-1/2\rangle \rangle = \\
&Z_{-1/2}-Z^{c.c.}_{-1/2} =-2i\gamma_{-1/2}.
\end{split}
\end{equation}
Making the analogous calculations for the each term of the expression~\eqref{e29} in the limit $\Omega\rightarrow \infty$ I will obtain the result~\eqref{mark}. 

\setcounter{equation}{0}
\def\theequation{B.\arabic{equation}}
{\bf Appendix B. Non-Markovian term}\\
In this appendix I examine the details of the obtaining expression~\eqref{e2967}. 
The determination of the operator $C^{\nu}$~\eqref{e32} leads to  
\begin{equation}\label{d22}
D^{\nu} =(C^{\nu})^{\star}=\sum\limits_{\mu\neq\nu}P^{\nu}L_{I}\frac{1}{w^{\nu}-w^{\mu}-i\varepsilon_{\nu\mu}}P^{\mu},
\end{equation}
where the action $"\ast"$ corresponds to the "star" conjugation (section 4). Taking into account the determinations~\eqref{e1779},~\eqref{e1796},~\eqref{e296},~\eqref{e234},~\eqref{e345} and replacement~\eqref{change},
in the lower-order approximation for the operator $\exp(-i\vartheta^{\nu}_{C}t)$  (see~\eqref{ae9})
\begin{equation}\label{e3524}
\exp(-i\vartheta^{\nu}_{C}t)\approx\exp(-iw^{\nu}t)P^{\nu}
\end{equation} 
it is easy to obtain 
\begin{equation}\label{e352}
\begin{split}
&\sum\limits_{\nu\neq 0}\langle \langle -1/2; -1/2\mid C^{\nu}\exp(-i\vartheta^{\nu}_{C}t)P^{\nu}D^{\nu}P^{0}\rho(0)\rangle\rangle = \\
&\sum\limits_{\nu\neq 0}\frac{\langle \langle -1/2;-1/2\mid L_{I}P^{\nu}
\exp(-i\vartheta^{\nu}_{C}t)P^{\nu} L_{I}\mid -1/2 ;-1/2\rangle \rangle}{(w^{\nu}-i\varepsilon)^{2}}=\\
&\sum\limits_{\rho}\Bigr{(}\exp(-i(E_{\rho}-E_{-1/2})t)\frac{<-1/2\mid H_{I}\mid\rho><\rho\mid H_{I}\mid -1/2>}{(E_{\rho}-E_{-1/2}-i\varepsilon)^{2}} +\\
&\exp(-i(E_{-1/2}-E_{\rho})t)\frac{<\rho\mid H_{I}\mid -1/2><-1/2\mid H_{I}\mid \rho>}{(E_{-1/2}-E_{\rho}-i\varepsilon)^{2}}
\Bigl{)}=\\
&\int\limits_{-\infty}^{\infty} \Bigr{(}\exp(-i(E_{1/2}+\omega -E_{-1/2})t)\times \\
&\frac{<-1/2\mid H_{I}\mid 1/2,\omega >< 1/2,\omega\mid H_{I}\mid -1/2>}{(E_{1/2}+\omega -E_{-1/2}-i\varepsilon)^{2}}d\omega + c.c.\Bigl{)}=\\
&{\Bigr{(}\frac{eg_{p}H_{1}}{4m_{p}}\Bigl{)}}^{2}\Bigr{(}\int\limits_{-\infty}^{\infty}f(\omega)\frac{\exp(-i(\omega -\omega_{0})t)}
{(\omega - \omega_{0} - i\varepsilon)^{2}}d\omega + c.c.\Bigl{)}.
\end{split}
\end{equation} 
Here,  index $\rho $ corresponds to the state $\mid 1/2, \omega>$,  $E_{\rho} =E_{1/2}+\omega$, $\omega_{0}=E_{-1/2}-E_{1/2}$.
In~\eqref{e352} the sum corresponds to summation (integration) over all discrete (continuous) indices.

\setcounter{equation}{0}
\def\theequation{C.\arabic{equation}}
{\bf Appendix C. Calculation of the integral}\\
The calculation of the integral in the expression~\eqref{e476} will be carried out in the kinematic region, where the function $f(\omega)$ satisfies the condition
\begin{equation}\label{funct}
\int\limits_{0}^{\infty}f(\omega)d\omega = 1.
\end{equation}
In this case, the integration  can be realized in the interval $\omega \in \left[  0, \infty\right)$.\\
 Since, $t>0$, we have to deform the contour of integration in the lower half-plane~\cite{dec}. 
In this region the application of the Cauchy's Theorem is possible (see Figure 2)
\begin{equation}\label{e12}
\begin{split}
&\int\limits_{\Gamma_{R}}\exp(-izt)F(z)dz= \int\limits_{0}^{R}\exp(-i\omega t)F(\omega)d\omega + \\
&\int\limits_{\gamma_{R}}\exp(-izt)F(z)dz + \int\limits_{-iR}^{0}\exp(-izt)F(z)dz =0,
\end{split}
\end{equation}
where
\begin{equation}\label{e123}
F(z) = \frac{f(z)}
{(z - \omega_{0} )^{2}}.
\end{equation}
If $f(z)$ is the Lorentzian distribution~\eqref{e867} then,
\begin{equation}\label{e765}
\lim\limits_{R\rightarrow \infty}\int\limits_{\gamma_{R}}\exp(-izt)F(z)dz \rightarrow 0 .
\end{equation}
It leads to
\begin{equation}\label{908}
\int\limits_{0}^{\infty}\exp(-i\omega t)F(\omega)d\omega = - \int\limits_{-i\infty}^{0}\exp(-izt)F(z)dz.
\end{equation}
The replacement  $z=-iy$, where $y=\xi /t$ makes it possible to reduce our  integral to the form
\begin{equation}\label{e2678}
\begin{split}
&\int\limits_{0}^{\infty}f(\omega)\frac{\exp(-i(\omega -\omega_{0})t)}
{(\omega - \omega_{0} - i\varepsilon)^{2}}d\omega\rightarrow \\
&-i\int\limits_{0}^{\infty} dyf(-iy)\frac{\exp(-i(-iy)t)}{(-iy-\omega_{0})^{2}}\exp(i\omega_{0}t) =\\
&-i\frac{\exp(i\omega_{0}t)}{\pi \delta\omega_{0}^{2}t}\int\limits_{0}^{\infty}d\xi\frac{\exp(-\xi)}{\left( 1+b\left( 1+i\frac{\xi}{a}\right) ^{2}\right)   \left( 1+i\frac{\xi}{a}\right) ^{2}},
\end{split}
\end{equation}
where $a=\omega_{0}t$ and $b=(\omega_{0}/\delta)^{2}$. It allows us to evalute the obtained integral with the use of the "Mathematica"-program. 
This leads to the expression
\begin{equation}\label{e3456}
\begin{split}
&\int\limits_{0}^{\infty}\left( f(\omega)\frac{\exp(-i(\omega -\omega_{0})t)}
{(\omega - \omega_{0} - i\varepsilon)^{2}} + c.c.\right) d\omega = \\
&\frac{2}{\pi\delta \omega_{0}^{2}t}\Bigl{(} A(t)\sin(\omega_{0}t) + B(t)\cos(\omega_{0}t)\Bigr{)}.
\end{split}
\end{equation}
The coefficients $A(t)$, $B(t)$ in~\eqref{e3456} are determined by the expressions~\eqref{e4152},~\eqref{e51648}.

\newpage

\begin{figure}[h]
\includegraphics[height=76.5mm]{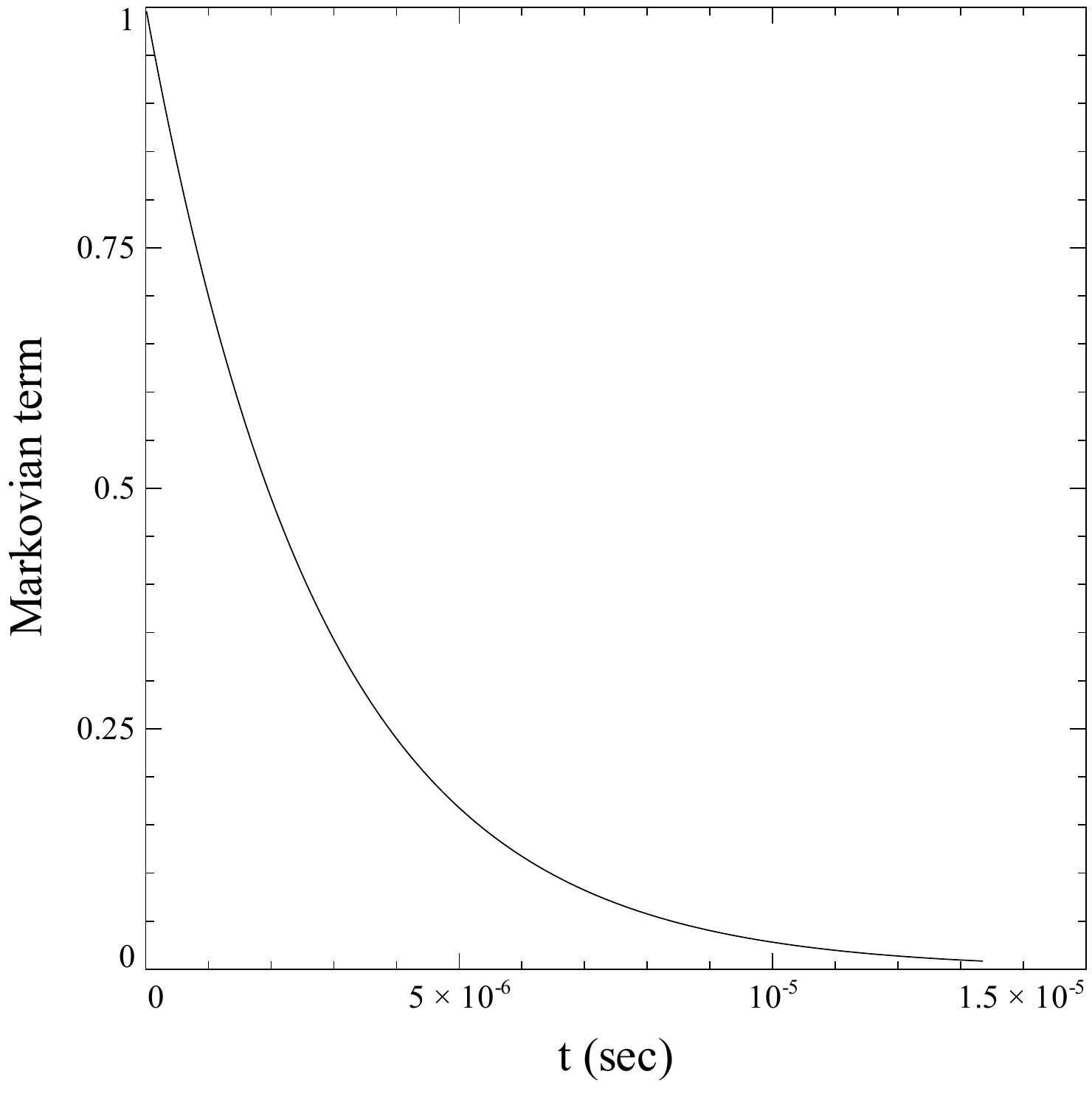}
\hspace{5mm}
\includegraphics[height=80mm]{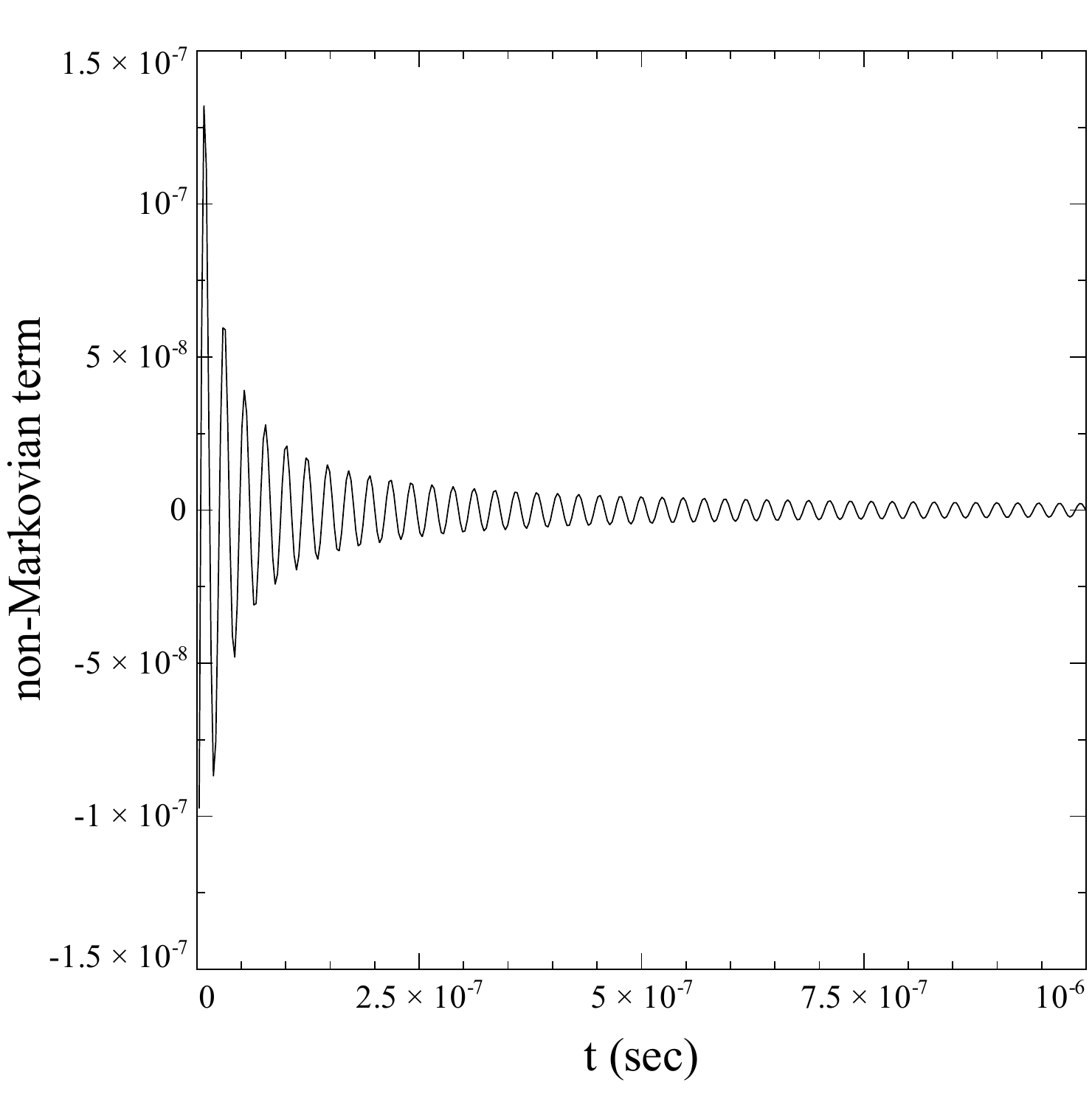}
\caption{}
\label{fig1}
\end{figure}

\begin{figure}[h]
\includegraphics[height=60mm~~~~~~~~~~~~~~~~]{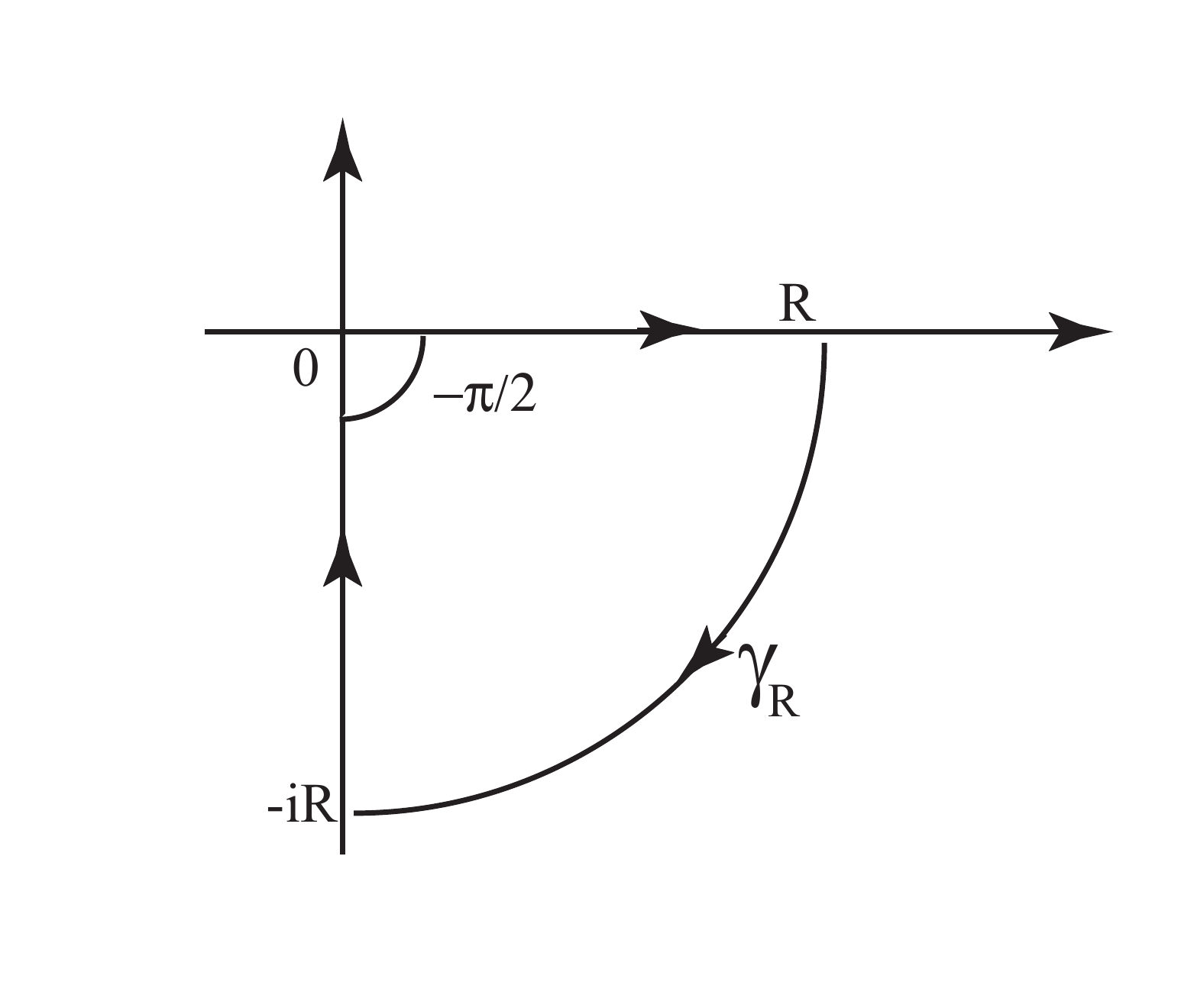}
\caption{}
\label{fig2}
\end{figure}

{\bf Figure Legends}

\vspace*{3mm} Fig.1: the Markovian and the non-Markovian terms; the proton, $H_{z}=10^{4}~Oe$, $H_{1}=10^{2}~Oe$, $\delta =0.033$ $m^{-1}$.

\vspace*{4mm} Fig.2:  a contour rotation.

\end{document}